\documentclass[12pt]{report}
\usepackage{fullpage}

\usepackage{epsfig}
\usepackage{cite}
\usepackage{graphicx}
\usepackage{latexsym}
\usepackage{amssymb}
\usepackage{amsmath}
\usepackage{amsfonts}
\usepackage{psfrag}
\usepackage{bm}
\usepackage{bbm}
\usepackage{subfigure}
\usepackage{epsfig}

\newcommand{\bt}{\begin{tabular}}
\newcommand{\et}{\end{tabular}}

\def\beq{\begin{eqnarray}}
\def\eeq{\end{eqnarray}}

\def\esp{\mathrm{e}}
\def\k{{\bf k}}
\def\d{{\rm{d}}}

\begin{document}

\pagenumbering{roman}

\begin{titlepage}

\thispagestyle{empty}

\begin{flushright}
{NYU-TH-09/012/16}
\end{flushright}
\vskip 0.9cm

\centerline{\Large \bf  Effective Field Theory for Quantum Liquid in  Dwarf Stars}      
              
\vskip 0.7cm
\centerline{\large Gregory Gabadadze$^1$ and Rachel A. Rosen$^2$}
\vskip 0.3cm
\centerline{\em $^1$Center for Cosmology and Particle Physics, 
Department of Physics}
\centerline{\em  New York University, New York, NY  10003, USA}
\vspace{0.3cm} 
\centerline{\em $^2$Oskar Klein Centre for Cosmoparticle Physics, Department of Physics}
\centerline{\em Stockholm University, AlbaNova SE-10691, Stockholm, Sweden}

\vspace{0.5cm}

\begin{center}
{\bf Abstract}
\end{center}

An effective field theory approach is used to describe quantum matter at greater-than-atomic but less-than-nuclear densities which are encountered in white dwarf stars.  We focus on the density and temperature regime for which charged spin-0 nuclei form an interacting charged Bose-Einstein condensate, while the neutralizing electrons form a degenerate fermi gas.  After a brief introductory review, we summarize distinctive properties of the charged condensate, such as a mass gap in the bosonic sector as well as gapless fermionic excitations.   Charged impurities placed in the condensate are screened with great efficiency, greater than in an equivalent uncondensed plasma.  We discuss a generalization of the Friedel potential which takes into account bosonic collective excitations in addition to the fermionic excitations.  We argue that the charged condensate could exist in helium-core white dwarf stars and discuss the evolution of these dwarfs.  Condensation would lead to a significantly faster rate of cooling than that of carbon- or oxygen-core dwarfs with crystallized cores.  This prediction can be tested observationally: signatures of charged condensation may have already been seen in the recently discovered sequence of helium-core dwarfs in the nearby globular cluster NGC 6397.  Sufficiently strong magnetic fields can penetrate the condensate within Abrikosov-like vortices.   We find approximate analytic vortex solutions and calculate the values of the lower and upper critical magnetic fields at which vortices are formed and destroyed respectively.  The lower critical field is within the range of fields observed in white dwarfs, but tends toward the higher end of this interval. This suggests that for a significant fraction of helium-core dwarfs, magnetic fields are entirely expelled within the core.


\end{titlepage}

\newpage

\tableofcontents
\addtocontents{toc}{\protect\thispagestyle{empty}}

\newpage

\pagenumbering{arabic}

\addcontentsline{toc}{chapter}{Introduction and Summary}
\chapter*{Introduction and Summary}

It is an everyday experience that by simply changing the temperature of a substance we can abruptly and dramatically change its macroscopic properties.  At normal earthly temperatures and pressures we observe, e.g., solids, liquids and gases, and the phase transitions between these states.  As we consider more extreme conditions out of the realm of our daily experience, say at temperatures near absolute zero or at the high densities that exist in the cores of many stars, phase transitions continue to occur and new states of matter emerge.  The principles of fundamental physics enable us to predict the properties of states of matter that have yet to be observed.  
   
At low temperatures and high densities, quantum mechanics becomes essential for describing the properties of a state of matter.  An example of this is the Bose-Einstein condensate.  Below a certain critical temperature the thermal de Broglie wavelengths of an ideal (or nearly ideal) gas of bosons will begin to overlap.  This critical temperature corresponds to when the quantum-mechanical uncertainties in the positions of the particles becomes greater than their inter-particle separation.  Below this temperature, the statistics governing a gas of indistinguishable bosons dictate that the particles will ``condense" into the same quantum state.  The first gaseous Bose-Einstein condensate was created in a laboratory seventy years after its existence was first predicted \cite{BEC}.  However, the extreme conditions required for the existence of such a quantum substance can occur outside the lab as well, in astrophysical objects.

The cores of white dwarf stars are composed of a particularly dense system of nuclei and electrons, with an average inter-particle separation much larger than the nuclear scale but much smaller than the atomic scale.  Because white dwarfs have exhausted their thermonuclear fuel, they evolve by cooling.  At high temperatures, the equilibrium state of the system is a plasma.  As the system cools below a certain temperature, the energy of the Coulomb interactions will significantly exceed the classical thermal energy.  Then, the standard classical theory holds that the nuclei will arrange themselves in such a way as to minimize their Coulomb energy, forming a crystal lattice  \cite{MestelRuderman}.  It is expected that in most white dwarfs, consisting of carbon, oxygen, or heavier elements, the crystallization transition takes place in the process of cooling (for a review, see, e.g., Ref. \cite{WDs}). 

In this work we will argue that for white dwarf stars with helium cores quantum effects become significant before the crystallization temperature is reached.  In these cases, the de Broglie wavelengths of the nuclei begin to overlap before crystallization can occur.  Then, because the helium-4 nuclei are bosons, the quantum-mechanical probabilistic ``attraction" forces the nuclei to undergo condensation into a zero-momentum macroscopic state of large occupation number.  The nuclei minimize their kinetic energy, while their collective fluctuations (phonons) have a mass gap. The majority of the phonons cannot be thermally excited since the phonon gap ends up being greater than the corresponding temperature. Therefore, after the phase transition all the thermal energy is stored in the near-the-fermi-surface gapless excitations of quasi-fermions.  We refer to this state as a charged condensate \cite{GGRR1,GGRR2,GGRR3,GGRR4,GGRR5,GGDP}. 

Because the repulsive Coulomb interactions between the ions dominate over the thermal energy of the system, the bosonic sector of the charged condensate is strongly coupled.  This is in contrast to the usual neutral Bose-Einstein condensate.  In this work we summarize an effective field theory approach to describing the charged condensate.  Within this framework we find properties of the charged condensate that are distinct from its neutral counterpart.  In particular, as mentioned above, we find that the spectrum of the collective bosonic excitations is gapped and the bosonic  contribution to the specific heat is exponentially suppressed at low temperatures.  Then, most of the entropy of the system is stored in the near-the-fermi-surface gapless fermionic excitations.

Furthermore, we find that electrically charged impurities in the condensate are screened to a high efficiency, more effectively than in an equivalent uncondensed plasma.  The static potential contains an exponentially suppressed term as well as a long-range oscillating piece.  The latter is due to gapless fermion excitations, and is similar to the Friedel potential.  However, the potential is  also suppressed due to an attractive phonon interaction, and we obtain an expression which has a long-range oscillatory nature but is highly suppressed compared to the conventional Friedel potential.

Such properties of the charged condensate have consequences for helium white dwarfs.  A condensed core dramatically affects the cooling history of the helium white dwarfs -- they cool faster than those with crystallized cores.  As a result, the luminosity function exhibits a sharp drop-off below the condensation temperature \cite{GGDP}.  Such a termination in the luminosity function may have already been observed in a sequence of the 24 helium-core white dwarf candidates found in the nearby globular cluster NGC 6397 \cite{24}. 

While the focus of this paper is the application of the charged condensate to helium white dwarfs, our methods are general and can be readily applied to other systems. The bosonic field can be generalized to any fundamental scalar field or a composite state, and the electromagnetic interaction could be replaced by any $U(1)$ abelian interaction.  The applications of a new state of matter are potentially diverse.

\vspace{0.3cm}

The structure of this work is as follows.  In chapter 1 we review the condensation of a gas of neutral bosons and give the expression for the standard critical temperature at which condensation occurs.  We review the condensation of a gas of weakly interacting bosons in the formalism of both a relativistic and a non-relativistic effective field theory.  

In chapter 2 we give arguments for the existence of the charged condensate.  A general mechanism for charged condensation in the context of a relativistic field theory is presented there and the spectrum of small perturbations above the condensate is determined.  Furthermore, it is shown in chapter 2 that electrically charged impurities are screened to a high degree due to an attractive phonon interaction.  There we consider the effects of fermion excitations on the electric potential and derive a generalized Friedel potential for the condensate.  We also briefly discuss the generalization of the Kohn-Luttinger potential.

In chapter 3 we argue that in helium-core white dwarf stars, the helium-4 nuclei may condense as they cool, instead of crystallizing.  A low-energy effective field theory description of the helium-4 charged condensate is developed, from which we recover the same characteristic properties that we found in the relativistic theory.  Furthermore, we consider the cooling rate of dwarf stars with condensed cores and show that the age of such dwarfs would be significantly shorter than those with crystallized cores.

In chapter 4 we look at the magnetic properties of the charged condensate. One would expect these to be similar to type II superconductors. Indeed, we find vortex-type solutions for magnetic flux tubes in the charged condensate.  From these we determine the magnitude of the external magnetic field for which it becomes energetically favorable to form vortices.  We discuss the applicability of the vortex solutions to magnetized helium-core white dwarfs, and  also consider the effect of a constant rotation on the magnetic field in the condensate of helium-4 nuclei. 

\vspace{0.3cm}

Let us make a few comments on the literature to emphasize the differences of the present approach.  The condensation of non-relativistic charged scalars has a long history, the original works being those by Schafroth \cite{Schafroth} in the context of superconductivity, and by Foldy \cite {Foldy} in a more general setup.  An almost-ideal Bose gas approximation was assumed in those studies.  For this assumption to be valid, densities had to be taken high enough to make the average inter-particle separation shorter than the  Bohr radius of a would-be {\it boson-antiboson} bound state \cite {Foldy}.  If the fermion number density is denoted by $J_0$,  and the mass of the scalar by $m_H$, this would be the case if $J_0^{1/3 }\gtrsim  \alpha_{\rm em} m_H$.  However, for a helium-electron system the above condition would translate into super-high densities, at which nuclear interactions would become significant.  Instead, in this work we study charged condensation in the opposite regime,  $J_0^{1/3} \ll \alpha_{\rm em} m_H$, where the nuclear forces play no role.  As a result, certain properties of the 
system -- such as important details of the spectrum  and the  screening of electric charge -- are  different.  Also, our method, which is based on symmetry and field theory principles, is different. 

In the context of a relativistic field theory the condensation of scalars was discussed in, e.g., Refs. \cite {Migdal,Linde,Kapusta}.  Pion condensation due to strong interactions is well known \cite{Migdal}.  In our work strong interactions play no role (except for providing the nuclei).  It was shown in Ref. \cite{Linde} that a constant background charge density strengthens spontaneous symmetry breaking when the symmetry is already broken by the usual Higgs-like nonlinear potential for the scalar.  In our work the scalar has a conventional positive-sign mass term and no Higgs-like potential.  The fact that the conventional-mass scalar could condense in a charged background was first shown in \cite{Kapusta},  in the case that the scalars have a nonzero chemical potential (see brief comments after eq. (4.6) in \cite {Kapusta}).  Moreover, a somewhat similar bosonic spectrum was already discussed in a work \cite{Lee2} in the context of superconductivity.  However, the dominant contributions to the thermodynamics of the charge condensate discussed here are due to near-the-fermi-surface electron excitations, which were absent in the the system considered in Refs. \cite{Kapusta,Lee2}.

The possibility of having a charged condensate in helium white dwarfs was previously pointed out in Ref.\cite{71}, where the condensation was studied using an approximate variational quantum-mechanical calculation in conjunction with numerical insights in a strongly-coupled regime of electromagnetic interactions.  The degree of reliability of such a scheme is hard to assess.  Furthermore, using the ordinary neutral Bose-Einstein (BE) condensation to describe the charged condensate, as is done in a number of works in the literature, is hard to justify. As we will see, properties of a neutral BE condensate differ significantly from those of the charged condensate considered here.  For instance, the specific heat at moderate temperatures in the former is due to a phonon gas, while in the latter it is due to the degenerate electrons.  We will discuss this feature of charged condensation and others in greater detail in what follows.

\vspace{0.1in}

The novel feature of our work is the development of an effective field theory approach to condensed matter at greater-than-atomic and less-than-nuclear densities, and its application to helium-core white dwarfs. This field-theoretic framework allows us to study the more subtle aspects of the condensate, including its spectrum and properties of its magnetic vortices.  The latter is especially hard to analyze without the field theory approach.  The generalization of the Friedel potential and the Kohn-Luttinger effect to a system with collective excitations of both bosonic and fermionic nature was obtained, using this method, in our work \cite{GGRR4}.  In addition, we are able to make concrete predictions as to the fast cooling of helium-core white dwarfs.

\vspace{0.1in}

In Ref. \cite{Dolgov,Dolgov2}, A. Dolgov, A. Lepidi, and G. Piccinelli have performed a one-loop calculation and found finite temperature effects in a general setup with condensed bosons.  These authors also obtained infrared modifications of the static potential for the scalar case, as we had earlier.  Our approach and that of Refs. \cite {Dolgov,Dolgov2} are complementary. We emphasize understanding the charged condensate in terms of effective field theory  and low-energy collective excitations. 

\vspace{0.1in}

Besides the first chapter which introduces the field theory description of a neutral condensate, the bulk of the present paper is based on our previous works on charged condensation \cite{GGRR1,GGRR2,GGRR3,GGRR4,GGRR5,GGDP}.  However, it also contains new technical and conceptual details that have not been published elsewhere.

\vspace{1.5cm}
\noindent {\bf Notations and Conventions:}  We work in natural units where $\hbar = c = k_B = 1$ unless explicitly stated otherwise.  The signature of the metric tensor taken to be $(+,-,-,-)$.  We use Heaviside-Lorentz units for Maxwell's equations.  Accordingly, the fine-structure constant is given by
\beq
\alpha_{\rm em} = \frac{e^2}{4 \pi} \simeq \frac{1}{137} \, .  \nonumber
\eeq

\chapter{The Neutral Condensate}
\label{ch1}

\section{Statistical mechanics of condensation}
At low temperatures and high densities, a gas of bosons exhibits fundamentally different behavior from a gas of fermions.  When the concentration of particles is sufficiently high so that the thermal de Broglie wavelengths of the particles begin to overlap, then quantum effects become important; the particles must be treated as truly indistinguishable with fermions obeying Fermi-Dirac statistics and bosons obeying Bose-Einstein statistics.  Because bosons do not obey the Pauli exclusion principle, they are free to occupy any state of the system in arbitrarily large numbers.  Thus at sufficiently low temperatures, the ground state of a system of bosons will be macroscopically occupied, forming a Bose-Einstein condensate.  In what follows we briefly review the statistics of a gas of bosons that lead to the critical temperature $T_c$ at which Bose-Einstein condensation occurs.

For a gas of bosons, the average number of particles in a state of energy $\epsilon$ is given by the usual Bose-Einstein distribution:
\beq
\label{BE}
N_{BE} (\epsilon,T)= \frac{1}{\esp^{(\epsilon-\mu)/T} -1} \, .
\eeq
Let us consider a non-relativistic gas of free bosons with energy $\epsilon = p^2/2m$, where $m$ is the mass of the particle.  Take the ground state of the system to have zero energy: $\epsilon_0 = 0$.  Then, as the non-relativistic chemical potential\footnote{The non-relativistic chemical potential is related to the relativistic chemical potential by $\mu_{NR} \equiv \mu-m$.} approaches zero $\mu_{NR} \to 0^-$, it is clear from the above expression that the number of particles in the ground state will become large:
\beq
\label{N0}
N_0 \rightarrow -\frac{T}{\mu_{NR}} ~~~~ {\rm as} ~~~~ \mu_{NR} \rightarrow 0^-\, .
\eeq
The total number of particles in the system is the sum of the number of particles in the ground state and in the excited states: $N_{tot} = N_0 + N_{ex}$.  In the $\mu_{NR} \to 0$ limit, one can calculate the number of particles that remain in excited states.  It is just the integral over all momentum states of excited particles:
\beq
\label{Num}
N_{ex} = V \int{\frac{d^3p}{(2\pi)^3} \frac{1}{\esp^{\, p^2/2mT} -1} }\, .
\eeq
Here $V$ is the volume of the system.  From this expression we can define a critical temperature $T_c$ at which the number of particles in excited states is equal to the total number of particles.  Above this temperature, there are negligibly few particles in the ground state and a nonzero (negative) chemical potential must be restored in expresion (\ref{Num}).  Below this critical temperature the ground state will be macroscopically occupied.  Taking $N_{ex} = N_{tot}$ and the density of particles to be $n = N_{tot}/V$ we find the usual critical temperature:
\beq
\label{Tcritical}
T_c = \frac{2 \pi}{m} \left(\frac{n}{\zeta(\tfrac{3}{2})} \right)^{2/3} \, .
\eeq
Here $\zeta$ is the Riemann zeta function: $\zeta(\tfrac{3}{2}) \approx 2.612$.  If we take the interparticle separation to be $d \equiv (\tfrac{4}{3} \pi n)^{-1/3}$ the critical temperature becomes
\beq
\label{Tcritical2}
T_c \simeq \frac{1.27}{m \, d^2} \, .
\eeq
This critical temperature corresponds to when the thermal de Broglie wavelengths of the free particles overlap with each other.  In other words, the condensation begins to occur when the quantum mechanical uncertainties in the positions of the particles become greater than the interparticle separation.  

The critical temperature is remarkable in that it can be significantly higher than the energy of the first excited state of the system.  Classically, we would expect that, in a system at temperature $T$, most particles would be in single-particle states with energy of order $T$, for arbitrarily small $T$.  However, the quantum statistics of a gas of bosons at low temperatures lead us to a strikingly different conclusion.  At low temperatures, but at temperatures still much higher than the energies of the lowest accessible excited states, bosons prefer to macroscopically occupy the ground state.

\section{The effective field theory}
For a large number of quanta that form a macroscopic state, the state can be adequately described in terms of an effective field theory of the order parameter, and its long wavelength fluctuations.  In particular, we will look for classical solutions of the equations of motion of the effective order-parameter Lagrangian.  How can a classical solution describe the condensate which is an inherently quantum phenomenon?  Denote the particle creation and annihilation  operators  by $a_0^\dagger$ and $a_0$ respectively; then the quantum-mechanical noncomutativity of these operators, $a_0^\dagger a_0 - a_0 a^\dagger_0 \sim \hbar$, becomes an insignificant effect of order ${\cal O}(\hbar/N)$, when the number of particles in the condensate state,  $\langle a_0^\dagger a_0 \rangle  \sim N$,  is large enough,  $N\gg 1$. Thus, the classical description of a coherent state with a large occupation number -- the condensate -- should be valid to a good accuracy \cite{LL}. On the other hand, collective excitations of the condensate itself should be quantized in a conventional manner.

The above arguments lead to the following decomposition of the order-parameter operator describing the condensate:
\beq
\Phi = \Phi_{cl} + \delta \Phi\,,
\label{classquant0}
\eeq
where $\Phi_{cl}$ denotes just a classical solution of the corresponding equations of motion, and describes the condensate of many zero-momentum particles, while $\delta \Phi$ should describe their 
collective fluctuations.  

In what follows we focus on the zero-temperature limit, even though realistic temperatures in, say, helium white dwarfs are well above zero (for calculations of the finite temperature effects, see \cite{Dolgov,Dolgov2}).  We will justify the validity of the zero-temperature approximation as we proceed (see, in particular, section 3.2).

\subsection{The relativistic EFT}
To describe the neutral condensate of bosons in terms of an effective field theory, we adopt the simplest Lagrangian for the order parameter $\phi$ that exhibits the condensation.  The field $\phi$ is a complex scalar field with a right sign mass term $m_H^2 > 0$ and a repulsive self-interaction with interaction strength $\lambda$:
\beq
\label{lagrN0}
{\cal{L}} = \vert \partial_{\mu} \phi \vert^2 - m_H^2 \phi^{\ast} \phi -\lambda (\phi^{\ast} \phi)^2  \,.
\eeq
This Lagrangian could contain higher order terms, however, they are generally suppressed by the short-distance cut-off of the theory and are irrelevant for our considerations.  Such a relativistic model of condensation was considered in Ref. \cite{Haber}.  The first microscopic theory of condensation 
in weakly interacting bose gases was developed by Bogoliubov \cite{bogo}.

For convenience, we switch notation and write $\phi$ in terms of its modulus and phase: $\phi = \tfrac{1}{\sqrt{2}} \sigma \esp^{i \alpha}$.  The Lagrangian (\ref{lagrN0}) becomes:
\beq
\label{lagrN}
{\cal{L}}= \frac{1}{2}(\partial_{\mu}\sigma)^2+\frac{1}{2}(\partial_\mu \alpha)^2 \, \sigma^2- 
\frac{1}{2} m_H^2 \, \sigma^2 - \frac{\lambda}{4} \sigma^4\, .
\eeq
Written in this form, it is evident that a nonzero value for $\partial_0 \alpha$ acts as a tachyonic mass for the scalar.\footnote{One could also introduce a chemical potential for the scalar which would have the same effect, however this term can be effectively absorbed into a redefinition of $\partial_0 \alpha$.}

Varying the Lagrangian w.r.t $\alpha$ gives the conservation of the scalar current density:
\beq
\label{eom2N}
\partial^\mu \left[(\partial_\mu \alpha) \, \sigma^2 \right] = 0.
\eeq
If we take $\partial_j \alpha = 0$, then the number density of scalars is constant in time.  Let us denote this number density by $J_0 =\partial_0 \alpha \, \sigma^2$.

Varying the Lagrangian with respect to $\sigma$ gives the following equation of motion:
\beq
\label{eom1N}
\Box  \sigma = [(\partial_\mu \alpha)^2-m_H^2] \, \sigma -\lambda \sigma^3\, .
\eeq
This equation admits a static solution $\bar{\sigma}$ that satisfies
\beq
\label{eom11N}
\frac{J_0^2}{\bar{\sigma}^3}-m_H^2 \bar{\sigma} -\lambda \bar{\sigma}^3 = 0\, .
\eeq
Let us take the quartic coupling $\lambda$ to be small.  In particular we assume $\lambda \ll m_H^3/J_0$.  Then the static solution for $\bar{\sigma}$, to first order in $\lambda$ is
\beq
\label{solN}
\bar{\sigma}  \simeq \sqrt{\frac{J_0}{m_H}} \left(1-\frac{\lambda}{4} \, \frac{J_0}{m_H^3} \right)\,.
\eeq
The nonzero vacuum expectation values (VEV) for $\sigma$
and $\partial_0\alpha$  imply  that the scalars are in the condensate phase, with nonzero number density: $J_0 = \partial_0 \alpha \, \sigma^2 \neq 0$.

From the Lagrangian (\ref{lagrN}) we can calculate the spectrum and propagation of perturbations above the condensate.  We find a heavy mode of mass $2 m_H$ which we ignore as it is beyond the scope of the low energy theory (since we assume that 
$m_H\gg J_0^{1/3}$), as well as a light mode.  The dispersion relation for the light mode is as follows:
\beq
\label{dispN}
\omega^2 \simeq \frac{\k^4}{4 m_H^2}+\frac{\lambda J_0}{2 m_H^3} \, \k^2 \, .
\eeq
Here we have taken the limits $\lambda J_0 \ll m_H^3$ and $\k^2 \ll m_H^2$.  In the absence of the self-interation term, i.e. when $\lambda \to 0$, this dispersion reduces to that for a free particle: $\omega = \k^2/2m_H$.  

The presence of the self-interaction term gives rise to the superfluidity of the neutral gas.  The long-wavelenth modes obey a linear dispersion relation:
\beq
\label{dispNlong}
\omega \simeq \sqrt{\frac{\lambda J_0}{2 m_H^3}} \, |\k | \,.
\eeq
The group velocity is then
\beq
\label{vgroupN}
v_{\rm gr} \simeq \sqrt{\frac{\lambda J_0}{2 m_H^3}}  \,.
\eeq
Such a linear dispersion relation (\ref{dispNlong}) meets the Landau criterion for superfluidity  \cite{Landau}.  At velocities less than $v_{\rm gr}$ the system experiences no loss of energy due to its motion.  Note that if the self-interaction coupling $\lambda$ were zero, then $v_{\rm gr}$ would also be zero and there would be no velocity at which the system could sustain a non-dissipative flow.  Self-interactions are essential to superfluidity.
  
The internal energy, specific heat and other thermodynamic quantities of the interacting condensate also follow from the above dispersion relations.  Using expression (\ref{dispNlong}) as the energy $\epsilon$  in the Bose-Einstein distribution (\ref{BE}), one can easily find the temperature dependence of the energy density of the gas of phonons ($U \propto T^4$) and of the specific heat ($C_V \propto T^3$).

\subsection{The non-relativistic EFT}
The relativistic effective Lagrangian adopted in the previous section is not necessarily the most appropriate description of the low energy condensate of non-relativistic particles, although captures many of its significant features.  It is overly restrictive in that it enforces Lorentz invariance- a symmetry we do not expect the low energy system to preserve.  
In this section we discuss a non-relativistic effective Lagrangian description of the neutral condensate.  We will see that in this formalism the condensate retains the distinctive features found in the relativistic theory, namely the equivalent dispersion relation for the light mode.

A non-relativistic effective order parameter Largangian that is consistent with the symmetries of the physical system can be written as:
\beq
\label{LgwwN}
{\cal L}_{eff} = {i\over 2} ( \Phi^*  \partial_0 \Phi -  (\partial_0 \Phi)^* \Phi)-{| \partial_j  \Phi|^2  \over 2m_H} -\frac{\lambda (\Phi^* \Phi)^2}{4m_H^2} \,,
\eeq
Again, higher order terms can be included in this Lagrangian, however they are irrelevant for our discussions. The possible quadratic term, $\mu_{NR}\Phi^*\Phi$,  
can be absorbed into  the first  term in (\ref {LgwwN}) by redefinition of the phase of the scalar field.
 We switch notation, writing $\Phi$ in terms of its modulus and phase $\Phi = \Sigma\, {\rm exp}(i\Gamma)$.  Written in terms of fields $\Sigma$ and $\Gamma$, the effective Lagrangian (\ref{LgwwN}) takes the following form: 
\beq
\label{Lgww1N}
{\cal L}_{eff} = -\partial_0 \Gamma \, \Sigma^2 - \frac{ (\nabla_j  \Sigma )^2}{2 m_H} -\frac{(\partial_j \Gamma)^2 \Sigma^2}{2 m_H} -\frac{\lambda \Sigma^4}{4 m_H^2}\,.
\eeq
Varying w.r.t. $\Sigma$ and $\Gamma$ gives the following equations of motion:
\beq
\label{eomN22}
\partial_0 (\Sigma^2) & = & - \frac{\partial^j (\Sigma^2 \partial_j \Gamma)}{m_H} \, , \\
\label{eomN21}
\nabla^2 \Sigma & = & 2 m_H \Sigma \, \partial_0 \Gamma+\Sigma \, (\partial_j \Gamma)^2+\frac{\lambda \Sigma^3}{m_H}  \, .
\eeq
Taking $\partial_j \Gamma = 0$, the first equation (\ref{eomN22}) gives the conservation of the scalar number density: $ \partial_0(\Sigma^2) = 0$.  Accordingly, we fix $\Sigma^2 = J_0$.  When $\lambda$ is nonzero, equation (\ref{eomN21}) has the following nonzero solution for $\partial_0 \Gamma$:
\beq
\label{solN2}
\partial_0 \Gamma = -\frac{\lambda J_0}{2 m_H^2} \, .
\eeq

From the Lagrangian (\ref{Lgww1N}) we find the dispersion relation for the scalar perturbation above the condensate:
\beq
\label{dispN2}
\omega^2 = \frac{\k^4}{4 m_H^2}+\frac{\lambda J_0}{2 m_H^3} \, \k^2 \, .
\eeq
This expression is identical to the approximate relation found for the light mode in the relativistic theory (\ref{dispN}).

\chapter{The Charged Condensate}
\label{ch2}

\section{A description of charged condensation}
Consider a neutral system of a large number of nuclei each having charge $Z$, and neutralizing electrons.  If the average inter-particle separation in this system is much smaller than the atomic scale, $\sim 10^{-8}$ cm, while being much larger than the nuclear scale,  $ \sim 10^{-13}$ cm, neither atomic nor nuclear effects will play a significant role.  Moreover, the nuclei can also be treated as point-like particles.   In what follows we focus on spin-0 nuclei with $Z \leq 8$ (helium, carbon, oxygen), and consider the electron number-density in the interval $J_0\simeq (0.1-5 {\rm ~MeV})^3$.  Then the electron Fermi energy will exceed the electron-electron and electron-nucleus Coulomb interaction energy.  Moreover, at temperatures below $\sim 10^7$ K, which are of interest here, the system of electrons form a degenerate Fermi gas.  

Since the nuclei (we also call them ions below) are heavier than the electrons, the temperature at which they'll start to exhibit quantum properties will be lower.  Let us define the ``critical'' temperature  $T_c$, at which the de Broglie wavelengths of the ions begin to overlap
\beq
\label{Tcrit}
T_c \simeq \frac{4\pi^2}{3m_Hd^2}\,,~~~~~d \equiv \left ( 3 Z \over 4 \pi   J_0 \right )^{1/3}\,,
\eeq
where $m_H$ denotes the mass of the ion (the subscript ``$H$'' stands for heavy), and $d$ denotes the average separation  
between the ions\footnote{The de Broglie wavelength above is defined as $\lambda_{dB}=2\pi/|\k|$, where $\k^2/2m_H = 3k_B T/2$. We define $T_c$ as the temperature at which $\lambda_{dB}\simeq d$.  Note that this differs by a numerical factor of $\sqrt{2\pi/3}$ from the standard definition of the {\it thermal} de Broglie  wavelength, $\Lambda \equiv \sqrt{ 2\pi /m_HT}$, that appears in the partition function of an ideal gas 
of number-density $n$ in the dimensionless combination $\Lambda^3 n$.}.

Somewhat below $T_c$ quantum-mechanical uncertainties in the ion positions become greater than the average inter-ion separation. Hence the latter concept loses its meaning as a microscopic characteristic of the system; the ions enter a quantum-mechanical regime of indistinguishability.  Then, the many-body wavefunction of the spin-0 ions should be symmetrized.  This unavoidably leads to a probabilistic ``attraction'' of the bosons to condense, i.e., to occupy one and the same quantum state.  We refer to the system of condensed nuclei and electrons as a charged condensate.

In the condensate the scalars occupy a quantum state with zero momentum.  Moreover, as we will show in section 2.3, small fluctuations of the bosonic sector have a mass gap, $m_\gamma = (Z e^2 J_0/m_H)^{1/2}$, which exceeds $T_c$ by more than an order of magnitude.  Therefore, once the bosons are in the charged condensate, their phonons cannot be thermally excited.  However, the gapless fermionic degrees of freedom near the fermi surface are thermally excited, and carry most of the entropy of the entire system \cite{GGRR1,GGRR2,GGRR3,GGRR4,GGDP}.

For further discussion it is useful to rewrite the expression for $T_c$ in terms of the mass density $\rho \equiv m_HJ_0$ measured in ${\rm g/cm}^3$:
\beq
\label{Tcrho}
T_c =  \rho^{2/3} \, \left ( {3.5 \cdot 10^2\over Z^{5/3}}\right )~K\,,
\eeq
where the baryon number of an ion was assumed to equal twice the number of protons, $A=2Z$ (true for helium, carbon, oxygen). 
Thus, for $\rho =10^6 {\rm ~g/cm}^3$ and helium-4 nuclei we get $T_c \simeq 10^6$ K, while for the carbon nuclei with the same mass density we get $T_c \simeq 2\cdot 10^5$ K. 

The temperature at which the condensation phase transition takes place, $T_{condens}$, is expect to be close to $T_{c}$ but
need not coincide with $T_{c}$.  Calculation of $T_{condens}$ from the fundamental principles of this theory is difficult.  However, we can obtain an interval in which  $T_{condens}$ should exist.  For this we introduce the following parametrization:
\beq
\label{Tcond}
T_{condens}= \zeta\, T_c\,,
\eeq
where $\zeta$ is an unknown dimensionless parameter that should depend on density more mildly than $T_c$ does.  Numerically, this parameter should lie in the interval $0.1 \ll  \zeta \lesssim 1$:  the point $\zeta=0.1$ corresponds to the Bose-Einstein (BE) condensation temperature of a free gas for which $T^{BE}_{condens}\simeq 1.3/m_Hd^2$ is known from fundamental principles (see chapter 1).  The condensation temperature in our system should be higher than $T^{BE}_{condens}$ since the repulsive interactions between bosons makes it easier for the condensation to take place \cite {Huang}.  In our case, these repulsive interactions are {\it strong} -- the Coulomb energy is at least an order of magnitude greater than the thermal energy in the system.  Hence, we would expect $\zeta \gg 0.1$.  Moreover, we will show in section 2.3 that the quantum dynamics of the fermions screen the mass of the bosons, in effect lowering $m_H$.  Thus the critical temperature defined by equation (\ref{Tcrit}) can be greater due to this smaller effective mass.  In what follows we will retain $\zeta $ in our expressions, but use the  value, $\zeta \simeq 1$, when it comes to numerical estimates. Perhaps a more accurate estimate of the condensation temperature can be obtained along the lines of \cite {Shuryak}.

\vspace{0.3cm}

The condensation will take place after gradual cooling only if $T_{condens}$ is greater than the temperature at which the substance would crystallize.  A classical plasma crystallizes when the Coulomb energy becomes about $\sim 180$  times greater than the average thermal energy per particle \cite{vanHorn,Ichi,DeWitt}.   This gives the following crystallization temperature\footnote{See chapter 3 for further discussion.}
\beq
\label{Tcryst}
T_{\text{cryst}}\simeq \rho^{1/3} \left 
(0.8\cdot 10^3 Z^{5/3} \right )~{\rm K}\,.
\eeq
Note that the density dependence of $T_c$ is different from that of $T_{\text{cryst}}$ --  for higher densities $T_c$ grows faster, making condensation more and more favorable!  One can define the ``equality'' density for which  $T_{condens}=T_{\text{cryst}}$:
\beq
\label{equality}
\rho_{\rm eq}= \left ( {2.3\over \zeta}\right )^3 Z^{10} ~{\rm g/cm}^3\,.
\eeq
For helium ($Z=2$) we find $\rho_{\rm eq}\simeq 10^4~{\rm g/cm}^3$, while for carbon ($Z=6$) we find $\rho_{\rm eq}\simeq  10^9~{\rm g/cm}^3$ (as mentioned above, we use $\zeta \simeq 1$). These results are very sensitive to the value of $\zeta$;  for instance, $\rho_{\rm eq}$ could be an order of magnitude higher if $\zeta \simeq 0.5$.  Regardless of this uncertainty, however, the obtained densities are in the ballpark of 
average densities present in helium-core white dwarf stars $\sim 10^6~{\rm g/cm}^3$.  For carbon dwarfs, they're closer to those expected in high 
density regions only \cite {GGDP}.

\vspace{0.3cm}
\begin{figure}[h!]
\begin{center}
\subfigure[Helium nuclei]{\epsfig{figure=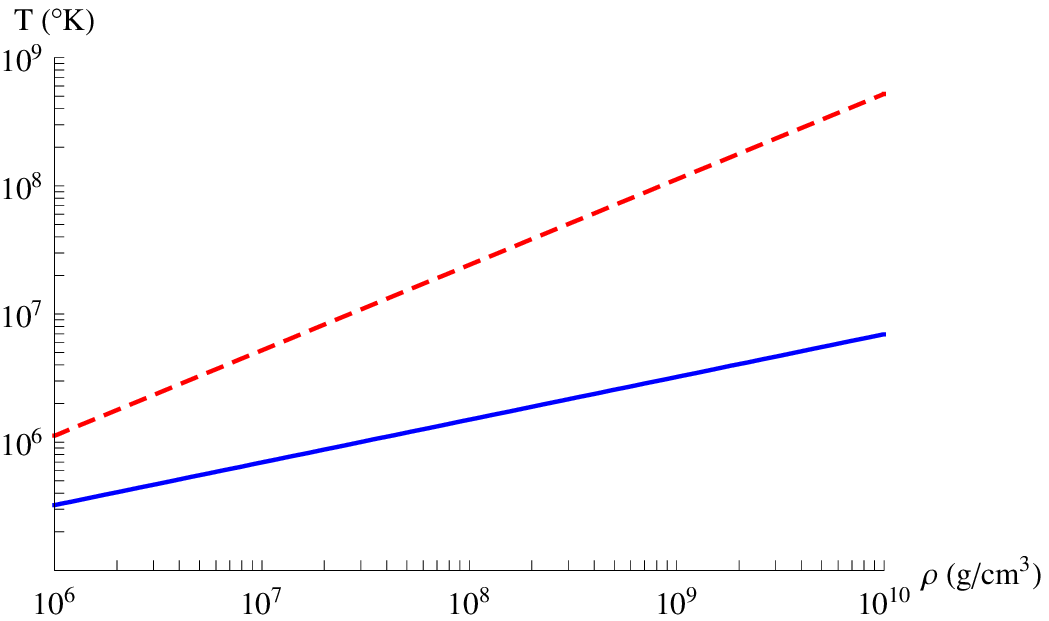,width=.45\textwidth}}
\hskip .15in
\subfigure[Carbon nuclei]{\epsfig{figure=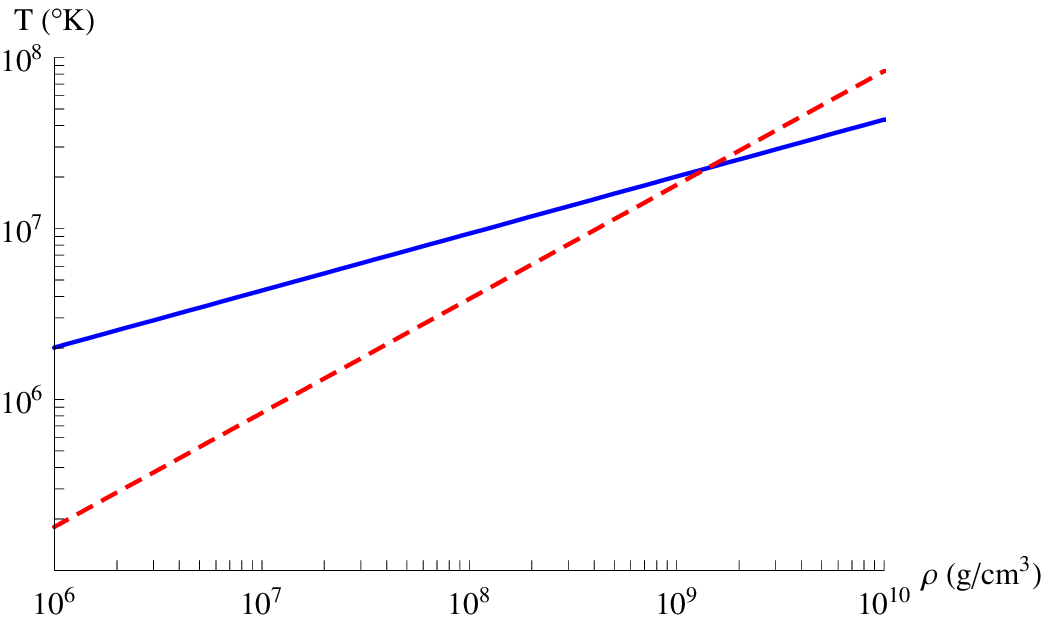,width=.45\textwidth}}\\
\end{center}
\caption{Crystallization temperatures (blue, solid line) and condensation temperatures (red, dashed line) as a function of mass density for systems of helium and carbon nuclei.}
\label{fig0}
\end{figure}

In Fig. \ref{fig0} we plot the the crystallization temperature and condensation temperature as a function of mass density for systems of both helium and carbon nuclei, taking $\zeta = 1$.  The solid line indicates the crystallization temperature and the dashed line the condensation temperature.  Typical core densities in white dwarf stars lie in the interval presented above, though the densities of most fall near the lower edge of this interval.  We see that for helium-core white dwarfs the condensation temperature is significantly greater than the crystallization temperature for all given densities.  The above considerations lead us to conclude that the dense system of the helium-4 nuclei and electrons may not solidify, but should condense instead.  We will return to this system in chapter 3.

Fig. \ref{fig0} also shows that for a carbon-core white dwarf, at typical densities, the critical temperature does not exceed the crystallization temperature and thus we would expect the nuclei to crystallize as usual.  
It can be seen that for superdense carbon white dwarfs with central density $\rho \simeq 10^{10}$ g/cm$^3$ the condensation temperature can exceed the crystallization temperature and thus these cores may also undergo condensation, however, this density is close to the neutronization threshold, as well as to the threshold where the relativistic  gravitational instability would set in, so the existence of dwarfs with such a high average density is questionable. On the other hand, such densities may still exist in small regions in the very core of 
dwarf stars; some effects of this were studied in \cite {GGDP}.  For oxygen-core white dwarfs (not shown) the crystallization temperature is always greater than the condensation temperature for relevant densities and thus we do not expect condensation to occur.

\vspace{0.3cm}

Is the charged condensate the ground state of the system at hand?  For the higher values of the density interval considered, the crystal would 
not exist due to strong zero-point oscillations.  At lower densities, the crystalline state has lower free energy (at least near zero temperature) due to more favorable Coulomb binding.  Hence, the condensate can only be a metastable state.  The question arises whether after condensation at  $\sim T_{condens}$ the system could transition at lower temperatures $\sim T_{\rm cryst}$ to the crystal state, as soon as the latter becomes available.

In the condensate, the boson positions are entirely uncertain while their momenta are equal to zero.  In order for such a system to crystallize later on, each of the bosons should acquire the energy of the zero-point oscillations of the crystal ions.  As long as this energy, $\sim (Z e^2 J_0/m_H)^{1/2}$, is much greater than $T_{\rm cryst}$, no thermal fluctuations can excite the condensed bosons to transition to the crystalline state. The latter condition is well-satisfied for all the densities considered in this work.  There could, however, exist a  spontaneous transition of a region of size $R_c$ to the crystallized state via tunneling.  The value of $R_c$, and the rate of this transition, will be determined, among other things, by 
tension of the interface between the condensate and crystal state, which is difficult to evaluate.  However, for an estimate, the following qualitative arguments should suffice: the height of the barrier for each particle is $(Z e^2 J_0/m_H)^{1/2}=m_\gamma$, while the number of bosons in the $R_c$ region $\sim R_c^3 J_0/Z$.  Hence, the transition rate should scale as ${\rm exp}(-m_\gamma J_0  R^4_c/Z)$.  Since we expect that  $R_c> 1/m_\gamma$, the rate is strongly suppressed for the parameters at hand.

\section{The relativistic EFT}
To described the charged condensate in terms of an effective field theory, we start by considering the simplest model that exhibits the main phenomenon: a generic, highly dense system of charged, massive scalars and oppositely charged fermions at zero temperature and infinite volume.  The scalar field could be a fundamental field or a composite state in the regime that compositeness does not matter.  The gauge field could be a photon or any other $U(1)$ field.

The scalar condensate is described by the order parameter $\phi$.  A nonzero vacuum expectation value (VEV) of $\phi$ implies that the scalars are in the condensate phase.  We adopt a relativistic Lorentz-invariant Lagrangian which contains the charged scalar field $\phi$ with right-sign mass term $m_H^2 > 0$, the gauge field $A_\mu$, and fermions $\Psi^\dagger,\Psi$ with mass $m_F$:
\beq
\label{lagr0}
{\cal{L}} = -\tfrac{1}{4}F_{\mu\nu}^2 + \vert D_{\mu} \phi \vert^2 - m_H^2 \phi^{\ast} \phi +
{\bar \Psi}(i\gamma^\mu D_\mu -m_F)\Psi + \mu_F \Psi^\dagger\Psi \,.
\eeq
The covariant derivatives in (\ref {lagr0}) are defined as  $\partial_\mu -ig_{\phi}A_\mu$ for the scalars, and  $\partial_\mu -ig_{\psi}A_\mu $ for the fermions. Their respective charges, $g_{\phi}$ and $g_{\Psi}$, are in general different.  For simplicity we take $g \equiv g_{\phi}=-g_{\Psi}$.

The chemical potential $\mu_F$ is introduced for the global fermion number carried by ${\Psi}$, for example lepton or baryon number.  The fermions in (\ref {lagr0}) obey the conventional Dirac equation with a nonzero chemical potential.  In particular, a self-consistent solution of the equations of motion implies that 
\beq
\label{muF}
\mu_F =  \sqrt{k_F^2+ m_F^2} + gA_0  \, , 
\eeq
where $k_F$ denotes the Fermi momentum of the background fermion sea.  The Fermi momentum is related to the number density of fermions $J_0$ as follows: $k_F = (3 \pi^2 J_0)^{1/3}$.  A nonzero chemical potential implies a net fermion number density in the system.  Since the fermions are electrically charged, they set a background electric charge density.  Such charged fermions would repel each other.  In our case, however, the fermionic charge will be compensated by the oppositely charged scalar condensate, as we will show below.  At distance scales that are greater than the average separation between the fermions their spatial distribution can be assumed to be uniform.  Then, the background charge density due to the fermions can be approximated as $J_\mu \equiv \bar{\psi} \gamma_\mu \psi = J_0 \delta _{\mu 0}$, where $J_0$ is a constant, fixed by (\ref{muF}).  One should take into consideration effects due to fermion fluctuations above $J_0$.  For now, however, we will assume that the fermions are frozen in ``by hand" and thus the averaging procedure is valid.  In later sections we will consider effects due to the 
dynamics of the fermions including their quantum loops.

Because the system also has a conserved scalar current, we can associate with it a chemical potential $\mu_s$.  For the Hamiltonian density, the inclusion of a chemical potential for the scalars results in the shift ${\cal{H}} \rightarrow {\cal{H}}' = {\cal{H}}-\mu_s J_0 ^{\rm{scalar}} $, where $J_0 ^{\rm{scalar}} \equiv -i [(D_0 \phi)^* \phi-\phi^* (D_0 \phi)]$ is the time component of the conserved scalar current.  For the Lagrangian density this shift can be written as a shift in the covariant derivative for the scalar $D_\mu \rightarrow D_\mu' = D_\mu-i\mu_s \delta_{\mu0}$.  In what follows primed variables ${\cal{H}}'$, ${\cal{L}}'$ will refer to those variables which include a nonzero chemical potential for the scalars.

We have not included a quartic interaction term for the scalar $\lambda (\phi^* \phi)^2$ in the Lagrangian (\ref {lagr0}).  This term could be present, but it is straightforward to check that our results will not be affected as long as $\lambda J_0 \ll m_H^3$.  The VEV of the scalar is fixed by it being energetically favorable for the bulk of the 
condensate to be neutral.  

In general, the scalar field could have an additional Yukawa term, $q (\phi^*{\bar \psi}_1\Gamma {\psi}_2+ {\rm h.c.})$, where $q$ is a coupling, $\Gamma$ denotes either the ${\bf 1}$ or $i\gamma_5$ matrix depending on the spatial parity of $\phi$, and ${\psi}_{1,2}$ denote fermions with different $U(1)$ charges that render the Yukawa term gauge invariant.  One or both of these fermions could be setting the background charge density $J_0$.  A fermion condensate $J_0^{1,2} \equiv \langle {\bar \psi}_1 {\psi}_2+ {\rm h.c.}\rangle$, if non-zero, could act as a source for the scalar.  In order for this not to significantly change our results, the condition $q J_0^{1,2} \ll m^2_H \langle \phi \rangle$ should be met.\footnote{The Yukawa coupling would also lead to the new terms in the fermion mass matrix.  Depending on the specific situation, this may or may not impose additional constraints.}  Due to this Yukawa coupling the scalar $\phi$ can decay.  In order for the condensate phase to form in the first place, the ``condensation time'' $\langle \phi \rangle^{-1}$ must be shorter then the lifetime of the $\phi$.

In the case of a system of charged nuclei and electrons the Yukawa terms are forbidden by the gauge $U(1)$ symmetry, so we will not be discussing them here. 

In addition, the effective Lagrangian (\ref{lagr0}) could contain the dimension-5 operator $ \propto \phi^{\ast} \phi \, {\bar \Psi} \Psi$.  Such a term would renormalize the mass of the fermions.  However, as long as 
the coupling constant that multiplies this term is sufficiently small this effect will not be significant. This is the case when the VEV of 
$\phi$ is much smaller than  the UV cutoff scale by which the above 
dimension-5 operator is suppressed. The latter conditions is fulfilled 
in our case since  $J_0 \ll m_H^3$.

\vspace{0.3cm}

The complex order parameter $\phi$ can be written in terms of a modulus and a phase $\phi = \tfrac{1}{\sqrt{2}} \sigma\, \esp^{i \alpha}$.  As per the discussion above, we treat the fermions as a fixed background density which couples to the gauge field as $-g A_{\mu} J^{\mu}$.  In terms of these variables the Lagrangian density becomes 
\beq
\label{lagr}
{\cal{L}}'=-\tfrac{1}{4} F_{\mu\nu}^2 + \tfrac{1}{2}(\partial_{\mu}\sigma)^2+
\tfrac{1}{2}(gA_\mu+\mu_s \delta_{\mu 0}-\partial_\mu \alpha)^2 \, \sigma^2- 
\tfrac{1}{2} m_H^2 \, \sigma^2 - g A_{\mu} J^{\mu} \, .
\eeq
From this form of the Lagrangian it is evident that a nonzero expectation value for $A_0$, or $\partial_0\alpha$,  or a nonzero chemical potential $\mu_s$ can give rise to a tachyonic mass for the scalars.  This effective tachyonic mass makes it possible for the scalar field to condense, as we shall now show.

Varying the Lagrangian with respect to $A_\mu$ and $\sigma$ gives the following equations of motion:
\beq
\label{eom1}
-\partial^\mu F_{\mu \nu}& =& g(gA_\nu+\mu_s \delta_{\nu 0}-\partial_\nu \alpha) \sigma^2-gJ_\nu \, ,  \\
\label{eom2}
\Box \, \sigma &= &[(gA_\mu+\mu_s \delta_{\mu 0}-\partial_\mu \alpha)^2-m_H^2] \, \sigma \, .
\eeq
The Bianchi identity for the first equation in (\ref{eom1}), can also be obtained by varying the action w.r.t. $\alpha$.  This gives the conservation of the scalar current:
\beq
\label{J0scalar}
\partial^\mu J_\mu^{\rm{scalar}} = \partial^\mu \left[(gA_\mu+\mu_s \delta_{\mu 0}-\partial_\mu \alpha) \, \sigma^2 \right] = 0.
\eeq

We can express the potential in terms of the gauge invariant variable $B_\mu \equiv A_\mu+\tfrac{1}{g} \partial_\mu \alpha$.  For a constant charge density, $J_\mu = J_0 \delta _{\mu 0}$, the theory admits a static solution: 
\beq
\label{EOMsol}
\langle g B_0 \rangle+\mu_s = m_H\,,~~~~~~~
\langle \sigma \rangle = \sqrt{\frac{J_0}{m_H}} \,.
\eeq
When $\sigma$ acquires a VEV, the gauge symmetry is spontaneously broken.   The mechanism of symmetry breaking for the charged condensate differs from the abelian Higgs model in that here the scalar field has a conventional positive-sign mass term.  Instead of giving a tachyonic mass to the scalars by hand, a nonzero expectation value for $B_0$ or a nonzero chemical potential $\mu_s$ act as a tachyonic mass term.  In particular, when $\langle g B_0 \rangle+\mu_s = m_H$, the scalar field condenses.  In the bulk of the condensate the scalar charge density exactly cancels the fermion charge density:  $J_0^{\rm scalar} = (\langle g B_0 \rangle+\mu_s) \langle \sigma^2 \rangle = J_0$.

\section{Spectrum of perturbations}
The uniform fermion background sets a preferred Lorentz frame.  We study the spectrum and propagation of perturbations in this background frame. For this we introduce small perturbations of gauge and scalar fields, $b_\mu$ and $\tau$, above their condensate values as follows:
\beq
\label{pert}
B_\mu(x)  = \frac{1}{g} (m_H-\mu_s) \delta_{\mu 0} + b_\mu(x)\,,~~~~\sigma(x) = \sqrt{\frac{J_0}{m_H}} + \tau(x)\,.
\eeq
In the quadratic approximation, the Lagrangian density for the perturbations reads
\beq
\label{Lpert}
{\cal L}_2= -\tfrac{1}{4} f_{\mu\nu}^2 + \tfrac{1}{2}(\partial_{\mu} \tau)^2+
\tfrac{1}{2}m_\gamma^2 b_\mu^2 + 2m_\gamma m_H \,b_0 \tau +...
\eeq
Here $f_{\mu\nu}$ denotes the field strength for $b_\mu$, and we have defined the following mass
\beq
\label{mgamma}
m_\gamma \equiv g \, \sqrt{\frac{J_0}{m_H}} \, .
\eeq
We have dropped all the fermionic terms as well as the cubic and quartic interaction terms of $b$'s and $\tau$.  This procedure is valid assuming that the perturbations are small compared to their condensate values, i.e. $g \, b_0 \ll m_H$ and $\tau \ll \sqrt{J_0/m_H}$.  The last term in (\ref{Lpert}) is Lorentz violating and is a consequence of having introduced the background fermion charge density.

It is also worth noting that the chemical potential for the scalars $\mu_s$ has disappeared from the Lagrangian (\ref{Lpert}).  In other words, the Lagrangian of perturbations is insensitive to whether the original theory (\ref{lagr}) contained a nonzero chemical potential for the scalars, or whether a nonzero VEV of the gauge invariant field $B_0$ 
is responsible for the condensation of the scalars.  Both cases 
give rise to the same spectrum of small perturbations.

Calculation of the spectrum of the theory is non-trivial but straightforward.  We briefly summarize the results.  First, $b_0$ is not a dynamical field, as it has no time derivatives in (\ref{Lpert}). Therefore, it can be integrated out through its equation of motion, leaving us with the equations for three polarizations of a massive vector $b_j$ ($j=1,2,3$), and one scalar $\tau$.  These constitute the four physical degrees of freedom of the theory.  The transverse part of the vector $b_j$ obeys the free equation 
\beq
\label{bj}
(\square +m_\gamma^2)b^T_j=0,~~{\rm where}~~~b^T_j \equiv b_j- {\partial_j\over \Delta}(\partial_k b_k)\,.
\eeq 
Therefore, the two states of the gauge field given by $b^T_j$ have mass $m_\gamma$.  Moreover, the frequency $\omega$ and the three-momentum vector $\k$ of these two states obey the conventional dispersion relation, $\omega^2 = m^2_\gamma+\k^2$. 

The longitudinal mode of the gauge field $b^L_j$, and the scalar $\tau$, on the other hand, give rise to the following Lorentz-violating dispersion relations (valid for $m_\gamma \neq 0$)
\beq
\label{LVdisp} 
\omega^2_{\pm} = \k^2+2m_H^2 + {1\over 2}m_\gamma^2 \pm
\sqrt{4\k^2m_H^2+ (2m_H^2- {1\over 2}m_\gamma^2 )^2} \,.
\eeq
The r.h.s. of (\ref{LVdisp}) is positive, thus the condensate background is stable w.r.t. small perturbations.  Both of these modes have masses which can be obtained by putting  $\k=0$.  For one mode this mass coincides with $m_\gamma$.  We refer to this mode as the longitudinal mode or the phonon, though in reality it is a linear combination of the scalar $\tau$ and the longitudinal gauge boson $b^L_j$.  The other mode has a mass $m_s=2m_H$ corresponding to the creation of a particle-antiparticle pair of scalars.  We refer to this as the scalar mode.  Interestingly, the group velocities of the transverse and longitudinal modes of the massive vector boson are different.  For $m_H \gg m_\gamma$, and for an arbitrary $\k$, the fastest ones are the transverse modes, they're followed by the scalar, and the longitudinal mode is the slowest.  All group velocities are subluminal.  

In the non-relativistic limit, for $m_H\gg m_\gamma$, the dispersion relations (\ref{LVdisp}) simplify to:
\beq
\label{pmdisp}
\omega^2_+ \simeq 4 m_H^2+2 \k^2 \, , ~~~~~~ \omega^2_- \simeq m_\gamma^2+\frac{\k^2 (\k^2-m_\gamma^2)}{4 m_H^2} \, , 
\eeq
The first of these corresponds to the creation of a particle-antiparticle pair.  The momentum term in this dispersion relation corresponds to the energy required for the propagation of one of these particles.  The second dispersion relation is more unusual.  To better understand it we consider the decoupling limit.  When electromagnetic interactions are turned off, $g \to 0$ and thus the mass $m_\gamma \to 0$.  Accordingly the second dispersion relation becomes $\omega_- = \k^2/(2 m_H)$.  This is exactly the energy required to propagate a {\it free} boson.  Thus the $m_\gamma$ terms in this dispersion relation can be thought of as a consequences of the electromagnetic interactions.  The mass gap $m_\gamma$ in the bosonic spectrum means that the Landau criterion is automatically satisfied and thus the bosons exhibit superfluidity. The dispersion relations for 
$\omega_{-}$ in (\ref {pmdisp}) also exhibits roton-like behavior 
(more on this in the  next section).

\vspace{0.3cm}

In our discussions so far we have treated the fermions as a fixed charge background $J_\mu  = J_0 \delta_{\mu 0}$.  We relax this assumption now and introduce dynamics for the fermions via the Thomas-Fermi (TF) approximation.  We consider the corrections to the spectrum of small perturbations due to these dynamics.  

The fermion number density is governed by the constant chemical potential $\mu_F$:
\beq
\label{muF2}
\mu_F = \sqrt{(3 \pi^2 J_0(x))^{2/3}+m_F^2}+g A_0(x) \, .
\eeq
Here we have related the local number density of fermions to the Fermi momentum via $J_0(x) = k_F(x)^3/(3 \pi^2)$.  In this way the number density of the fermions $J_0$ gets related to the electric potential $A_0$.  For relativistic fermions
\beq
\label{J0}
J_0(x) = \frac{1}{3 \pi^2} (\mu_F-g A_0(x))^3 \, .
\eeq
Consequently, fluctuations in $J_0(x)$ can be expressed in terms of fluctuations in the potential $b_0(x)$.  As a result, the coefficient in front of $b_0^2$ gets modified as compared to (\ref{Lpert})
\beq
\label{LpertF}
{\cal L}_2= -{1\over 4} f_{\mu\nu}^2 + \tfrac{1}{2}(\partial_{\mu} \tau)^2+
\tfrac{1}{2}m_0^2 b_0^2-\tfrac{1}{2}m_\gamma^2 b_j^2 + 2m_\gamma m_H \,b_0 \tau +...
\eeq
where ${m}_0^2 \equiv  m^2_\gamma+ {(g^2/ \pi^2) \, (3\pi^2 J_0)^{2/3} }$.  The latter term is simply the Debye mass squared.  The introduction of the fermion dynamics via the TF approximation breaks the degeneracy between the ``electric" and ``magnetic" masses of the gauge field.

Calculation of the spectrum is once again straightforward. The two transverse components of the gauge field are not affected by the addition of the fermion dynamics.  They still propagate according to the usual massive dispersion relation $\omega^2 = m^2_\gamma+\k^2 $.  The dispersion relations of the longitudinal and scalar modes (\ref{LVdisp}) become
\beq
\label{disp2}
\omega_{\pm}^2 &=& {\bf k}^2 \left ( {m_0^2 + m^2_\gamma \over 2 m_0^2 }
\right ) + {2M^4 \over m_0^2} + {m^2_\gamma \over 2 }  \nonumber  \\
& \pm &  \sqrt {4 {\bf k}^2 {M^4 m^2_\gamma \over m_0^4}+ \left [
{2M^4 \over m_0^2} - {m^2_\gamma \over 2}+ {\bf k}^2
\left ( {m_0^2 - m^2_\gamma \over 2 m_0^2} \right) \right]^2}\,.
\eeq
where $M \equiv \sqrt{m_H m_\gamma}$.  The r.h.s of (\ref{disp2}) is positive for arbitrary $\k$ and thus the charged condensate background is stable w.r.t. small perturbations.  All the group velocities obtained from (\ref{disp2}) are subluminal.

The solution with the minus subscript corresponds to the longitudinal component of the massive vector field, with $\omega_{-}^2 ({\bf k}=0)=m^2_\gamma$.  Though the dispersion relation is different with the introduction of the fermion dynamics, the mass gap for this mode remains the same.  The solution with the plus subscript corresponds to the scalar mode and its mass squared in this frame is $\omega_{+}^2 ({\bf k}=0)= 4M^4/m^2_0$.  Prior to introducing the fermion fluctuations, the mass of this mode was given by $\omega_{+}^2 ({\bf k}=0)= 4m_H^2=4M^4/m^2_\gamma$.  As $m_0^2$ is greater than $m_\gamma^2$ by definition, we see that the effect of the fermion dynamics is to lower the effective mass of the bosonic particle-antiparticle pair.  Hence  the effective mass of the condensed bosons get screened due to fermion quantum effects, suggesting  that 
the condensation phase transition temperature may actually be 
even higher than what we use here, and adopting $\zeta=1$ in chapter 
1 may be a conservative choice.   We cannot, however,  directly use the above expression for the effective mass since the effective theory used 
for its derivation is not expected to be reliable at the 
scale of the  mass itself. 
  
In the limit that the fermions are non-dynamical (i.e. are frozen ``by hand'' or some other dynamics), then $m_0 \to m_\gamma$, and the solutions reduce to the ones obtained in (\ref{LVdisp}).  However, for most physical setups we will find that the difference between $m_0$ and $m_\gamma$ is greater than $m_\gamma$.  Therefore, the fermion dynamics introduce an additional screening of the electrostatic interactions.  In fact, this is just the usual Debye screening.

\section{Screening of electric charge}
As a next step we'd like to discuss the screening of electrically charged impurities in the condensate.  To determine the screening length we consider a small, spherically symmetric object with a nonzero charge placed in the condensate.  Outside of the charge, in the condensate, the equations of motion for a static ${A_0}$ and $\sigma$ as derived from (\ref{eom1}) and (\ref{eom2}) are:
\beq
\label{eomstat}
-\nabla^2 A_0 +g (g A_0+\mu_s) \sigma^2 = g J_0 \,,~~~~~
-\nabla^2 \sigma = [(g A_0+\mu_s)^2 - m_H^2] \sigma \,.
\eeq
In terms of the perturbations of the fields above their condensate values (\ref{pert}), we can write these equations as
\beq
\label{eompert}
\nabla^2 b_0  = m_\gamma^2 b_0+2 M^2 \tau \,,~~~~~
-\nabla^2 \tau = 2 M^2 b_0 \,,
\eeq
where again we have defined $M \equiv \sqrt{m_H m_\gamma}$.  The above equations are valid as long as we focus on solutions that satisfy  $\tau \ll \sqrt{J_0/m_H} $ and $g \, b_0 \ll m_H$. 

The regime of physical interest is the one in which $m_H \gg J_0^{1/3}$.  This will be applicable to the system of helium-4 nuclei and electrons in helium white dwarf stars.  In this case $M \gg m_\gamma$, and we can neglect the first term on the r.h.s. of the equation for $b_0$ in (\ref{eompert}).  For large $r$ we require that $b_0, \, \tau \rightarrow 0$.  The boundary conditions select the decaying functions:
\beq
\label{b0sol}
b_0(r) &\simeq& \frac{\esp^{-Mr}}{r} \left[c_1 \sin(M r)+c_2 \cos(M r)\right] \,, \\
\label{tausol}
\tau(r) &\simeq&  \frac{\esp^{-Mr}}{r} \left[-c_1 \cos(M r) +c_2 \sin(M r)\right] \,.
\eeq
The constants $c_1$ and $c_2$ are to be determined by matching these solutions to those in the interior of the small, charged object.  Thus, for a probe particle, the screening occurs at scales greater than  
$1/M$.  When $m_H \gg J_0^{1/3}$, $1/M$ is shorter that the average inter-particle separation $d \propto J_0^{-1/3}$.  Although this strong screening may well be a reason why the condensation of charged bosons takes place in the first place, this statement needs some qualifications.  It may seem that the exponent ${\rm exp} (-Mr)$ is due to a state of mass $M$.  The distance scale $1/M$ is shorter than the average inter-particle separation -- an effective short-distance cutoff of the low-energy theory.  Then, a state of mass $M$, if existed, would have been beyond the scope of the low-energy field theory description, and the above potential would have been unreliable.  There is an explanation of this scale in terms of a cancellation between the potentials due to the two long-wavelength modes -- Coulomb and ``phonon''  quasiparticles --  both of which are much lighter than $1/d$, and are well-within the validity of the effective field theory.  
So the obtained result is well within the scope  of the effective field theory for $r> d$.  

To study this effect in more detail we start by calculating the gauge boson propagator.  For now we once again treat the fermions as ``frozen in:" $J_\mu = J_0 \delta_{\mu 0}$.  We will return to their dynamics at the end of this section.  The gauge boson propagator can be determined from the Lagrangian of small perturbations (\ref{Lpert}).  It is useful to integrate out the $\tau$ field. The remaining Lagrangian takes the form:
\beq
\label{Ltau}
{\cal L}_{2}= -{1\over 4} f_{\mu\nu}^2 +{1\over 2} m_\gamma^2 b_\mu^2 +
{1\over 2} \,b_0 {4M^2\over \square}b_0. 
\eeq
This Lagrangian contains four components of $b_\mu$, and no other fields. The first two terms in 
(\ref{Ltau}) are those of a usual massive photon with three degrees of freedom.  The last term is unusual, as it gives rise to the dynamics to the timelike component of the gauge field.  This term emerged due to the mixing of $b_0$ with the dynamical field $\tau$ in (\ref{Lpert}), and since we integrated out $\tau$, 
$b_0$ inherited its dynamics in a seemingly nonlocal way. 

This form of the Lagrangian (\ref{Ltau}) is useful for calculating the propagator.  Indeed, the inverse of the quadratic operator that appears in (\ref{Ltau}) has poles which describe all the four propagating 
degrees of freedom.  The full momentum-space propagator is given by
\beq
\label{Dprop}
\lefteqn{D_{\mu \nu} (p)= \frac{1}{p^2-m_\gamma^2} \times}  \\
&&  \left[ -g_{\mu \nu} +  \frac{4m_\gamma^2M^4 \delta_{\mu 0} \delta_{\nu 0}+(p^2 (p^2-m_\gamma^2)+4M^4)p_\mu p_\nu-4\omega M^4(\delta_{\mu 0} p_\nu+p_\mu \delta_{\nu 0})}{p^2 m_\gamma^2 (p^2-m_\gamma^2)-4M^4(\omega^2-m_\gamma^2)}\right] \, , \nonumber
\eeq
where $p$ is the four-momentum and $\omega = p_0$.  In the limit that $M \to 0$ (with fixed $m_\gamma$) this propagator describes a usual massive vector boson:
\beq
\label{DpropM}
D_{\mu \nu} \to \frac{1}{p^2-m_\gamma^2}  \, \left[ -g_{\mu \nu}+\frac{p_\mu p_\nu}{m_\gamma^2} \right] \, .
\eeq
For nonzero $M$ the propagator is modified by Lorentz-violating terms.

Sandwiched between two conserved currents $J_\mu$ and $J^\prime_\nu$,  the propagator takes the form:
\beq
\label{amplitude}
J^{\mu} D_{\mu \nu} J^{\nu \prime} = \frac{J_0  \left( 1-\frac{4 M^4 \omega^2}{p^2 m_\gamma^2(p^2-m_\gamma^2)} \right) J_0^\prime}
{ -p^2 +m_\gamma^2 + \left (1- {\omega^2\over m_\gamma^2} \right) {4M^4\over -p^2} }
- \frac{J_j J_j^\prime}{-p^2 +m_\gamma^2}   \,.
\eeq
The poles of this propagator describe two transverse photons with mass $m_\gamma$, one heavy mode with mass $2m_H$, and a light phonon with mass $m_\gamma$.  Their dispersion relations are those found in (\ref{bj}) and (\ref{LVdisp}).

In particular, we are interested in a static potential for a point source.  This can be obtained from the propagator (\ref{amplitude}):
\beq
\label{photonphonon}
V(\k) = D_{00}(\k, \omega=0)= \frac{1}{\k^2 +m_\gamma^2} - \frac{1}
{\k^2 +m_\gamma^2+ \k^2 (\k^2+m_\gamma^2)^2/4M^4}\,.
\eeq
The first term on the r.h.s. can be thought of as an repulsive screened Yukawa potential while the second term can be interpreted as an attractive potential due to a phonon.  This second term in (\ref {photonphonon}) has three poles.  The residue of one pole exactly cancels that of the first term of (\ref {photonphonon}).  The remaining two poles describe both the heavy state of mass $2 m_H$ which is unimportant for the low-energy physics, and a light state, which actually is the phonon.  It's the light mode found in (\ref{LVdisp}) that belongs to the spectrum of the low-energy effective field theory.  For simplicity of the discussions, we'll be using a somewhat imprecise language  by calling the whole second term in  (\ref{photonphonon}) the phonon contribution. 

The phonon in this case is a collective excitation of the motion of charged scalars within the fermion background.  The cancellation due to this light mode gives rise to the exponential  ${\rm exp} (-Mr)$, and not a hypothetical state of mass $M$.  At scales larger than $1/M$, which are of primary interest, the phonon potential cancels the gauge potential with a high accuracy.  This cancellation is reliable at scales that are much greater than $J_0^{-1/3}\gg M^{-1}$, and takes place already at scales that are much shorter that the photon Compton wavelength $m^{-1}_\gamma \gg J_0^{-1/3}$.

In a Lorentz-invariant theory having a negative sign in front of a propagator, such as the one in the second term of (\ref{photonphonon}), would suggest the presence of a ghost-like state.  However, this is not the case in a Lorentz-violating theory described by our Lagrangian (\ref{Lpert}).  The fact that there are no pathologies in (\ref{Lpert}), such as ghost and/or tachyons, can be seen by calculating the Hamiltonian density:
\beq
\label{ham2}
{\cal H}_2=  \frac{1}{4} f^2_{ij} +\frac{1}{2} \pi_j^2 + \frac{1}{2} P_\tau^2 + \frac{1}{2}(\partial_j\tau)^2+\frac{1}{2} m_\gamma^2 b_j^2 +\frac{1}{2 m_\gamma^2} (\partial_j \pi_j - 2 m_H m_\gamma \tau)^2 \,.
\eeq
Here,  $\pi_j \equiv -f_{0j}$ and $P_\tau\equiv \partial_0\tau$.  The Hamiltonian is positive semi-definite.  Hence, no ghosts or tachyons are present.  Moreover, consistent with ones expectation, the second term in (\ref {photonphonon}) disappears in the limit $M\to 0$,  where the Lorentz invariance of (\ref {Lpert}) is restored.

The static potential (\ref{photonphonon}) can be simplified:
\beq
\label{static}
V(\k) = \left ( \k^2 +m_\gamma^2 + {4M^4\over \k^2} \right )^{-1}\,.
\eeq
The first, second, and third terms on the r.h.s. of (\ref{static}) are due to the respective terms in (\ref{Ltau}).  Interestingly, when $m_H \gg m_\gamma$, there is no scale at which the photon mass term in  (\ref{static}) would dominate:  for $\k^2 \gtrsim m^2_\gamma$ the mass term is sub-dominant to the $\k^2$ term, while for $\k^2 \lesssim m^2_\gamma $ it is sub-dominant to the $M^4/\k^2$ term.  The term $M^4/\k^2$ is coming from the phonon cancellation and gives rise to significant modification of the propagator in the infrared.  

The above described properties of the propagator can be seen by calculating the coordinate space potential from (\ref{static}).  We find, in agreement with (\ref{b0sol}): 
\beq
\label{potential}
V(r) \equiv (Q_1e Q_2e) \int {d^3 \k \over (2\pi)^3} e^{i\k {\bf x}}G(\omega=0, \k) 
\propto  {Q_1Q_2 \alpha_{\rm em} e^{-Mr}\over \,r}{\rm cos} (Mr)\,,
\eeq
in which we assumed that  $r=|{\bf x}|\gg 1/M$ and $m_\gamma  \ll M$.  This potential is  sign-indefinite  and undergoes modulated oscillations between repulsion and attraction.  There are an infinite number of points in the position space where the force between classical  charges would vanish.  These are points where  
\beq
\label{po}  
{dV\over dr}(r=r_n)=0, ~~~~~~n=1,2,...
\eeq
Any two charged probe particles separated by a distance $r_n \gtrsim J_0^{-1/3}$, where our calculations are reliable, would stay in a static equilibrium as long as $V(r_n)<0$. However, the potential is too shallow and any realistic temperature effects would kick the probe charges 
out the potential wells in (\ref {po}).

Furthermore, the finite temperature corrections could  modify the properties of the condensate itself, however, in this particular case, 
due to a high mass gap,   the main properties of the condensate  should remain valid at temperatures well-below the condensation point.  For instance, in white dwarfs with temperature $10^6 - 10^7$ K we would expect the dominant  temperature-dependent corrections to  the potential to be proportional to $T/ J_0^{1/3} \sim (10^{-4}-10^{-3})\ll 1$, which are negligible.

Before proceeding to the next section we make two comments.  First, in the $m_H \to \infty$ limit  one would expect the heavy scalars to decouple.  It is not exactly clear from (\ref{photonphonon}) how such a decoupling takes place, and what is its interpretation.  In the limit $m_H \to \infty$, which implies $M \to \infty$,  the phonon effects should disappear. 
This certainly is  the case in the full amplitude discussed before. However,  taking this limit in (\ref{photonphonon}) (or in (\ref{static})) results in a vanishing of the whole potential.  This is an artifact of using the static approximation and can be understood in the following way: the phonon mixes with the timelike component of the gauge field, and because of this acquires an instantaneous part.  Then, the instantaneous parts in (\ref{photonphonon}) cancel between the gauge and photon contributions.  However, the dynamical part of the phonon also reduces to zero, as the group velocity of the phonon vanishes in the $m_H \to \infty$ limit.

This can be seen by looking at the dispersion relation for the phonon which was obtained in (\ref{pmdisp}).   For the relevant momenta $M^2 \gg \k^2$ the dispersion relation for $\omega_-$ gives the phonon group velocity
\beq
\label{vgr}
v_{\rm gr} \simeq  {m_\gamma |\k| (2\k^2 -m_\gamma^2)\over 4M^4}\,.
\eeq
This vanishes in the $m_H \to \infty$ limit. 

Note that for  $\k^2 \simeq m_\gamma^2/2$ the phonon group velocity also vanishes for finite $m_H$. This  describes a state of a nonzero momentum but zero group velocity.  The energy of this state is also nonzero, and to a good approximation equals to $m_\gamma$.  These properties are  similar to those of a roton in superfluid helium II.  Moreover, for excitations with $\k^2 > m_\gamma^2/2$ the group velocity is positive, while in the opposite case, $\k^2 < m_\gamma^2/2$, it becomes negative (i.e. the direction of 
the momentum and that of group velocity are opposite to each other).  These excitations resemble the positive and negative group velocity rotons in superfluid helium II.

The second comment concerns the limits of applicability of the linearized approximation.  The expansion   in (\ref{Ltau}) is only valid when the perturbations $\tau$ and $b_0$ are much smaller than their condensate values.  For scalar field this means that $\tau \ll \sqrt{J_0/m_H}$ and for the gauge field $g b_0 \ll m_H$.  In the limit $m_H\to \infty$, the domain of applicability of the linearized results shrinks to zero.  This suggests that the geometric size of the region in which one can meaningfully talk about  the charged condensate should be greater than a certain critical size that scales as $\sim  (\sqrt{J_0/m_H})^{-1}$.  The latter tends to infinity as $m_H\to \infty$.

\vspace{0.3cm}

In our treatment of the screening of electric charge we have treated the fermions as ``frozen in.''   Such an approximation would be physically justifiable if, say, the fermions were fixed in a crystal lattice.  However, in many physical circumstances, including the helium white dwarf system to be discussed later, this is not a good approximation.  The fermion fluctuations should be taken into account.  This was done in the previous section for the spectrum of small perturbations using the Thomas-Fermi (TF) approximation.  The result was that the timelike component of the gauge field acquired an contribution to its mass.  This mass term modifies the first of the equations of motion for small perturbations (\ref{eompert}):  
\beq
\label{eompert2}
\nabla^2 b_0  = m_0^2 b_0+2 M^2 \tau \,,~~~~~
-\nabla^2 \tau = 2 M^2 b_0 \, ,
\eeq
where $m_0^2$ is defined as above as the sum of the photon mass squared and the Debye mass squared.  The regime of physical interest, $m_H \gg J_0^{1/3}$, corresponds to $M \gg m_0$.  If we again calculate the potential outside a static probe charge using these modified equations, the $m_0$ term in the above expression for $b_0$ is subdominant compared to the $M$ term.  Thus the potential obtained in (\ref{b0sol}) for a probe charge is still valid with the inclusion of this additional mass term.

However, the TF approximation does not capture the significant property of the fermion system related 
to the possibility of exciting gapless modes near the Fermi surface.  We can incorporate these effects into our results by calculating the one-loop correction to the propagator (\ref{Dprop}). For this, we restore back in the Lagrangian (\ref{Ltau}) the fermion kinetic, mass and chemical potential terms and, upon calculating the gauge boson propagator, we will take into account the known one-loop gauge boson polarization diagram. This diagram is suppressed by an additional power of the electromagnetic coupling constant $\alpha_{\rm em}=e^2/4\pi$, and one would expect the quantum correction to be insignificant.   However, this is not the case for the following subtle reason.  The one-loop correction introduces branch cuts in the propagator, which give rise to additional contributions to the static potential in the position space.  These additional terms have oscillatory nature with a {\it power-like} decaying envelope.  Even though they are formally suppressed by $ {\cal O}(\alpha_{\rm e}^2)$, to a good approximation they end up being ${\cal O}(1)$, and can dominate over the exponentially suppressed term at sufficiently large distances.

The static potential obtained from the $\{00\}$ component of the propagator $ D_{00}$ with the one-loop correction is given by  
\beq
\label{V00}
V(k)\equiv D_{00}(\omega =0, \k) = \left ( \k^2 +m_\gamma^2 +  {4M^4\over \k^2} + F (k^2,k_F,m_F)  \right)^{-1}\,.
\eeq
The function $F (k^2,k_F,m_F)$ is due to the one-loop photon polarization diagram and includes both the vacuum and fermion matter contributions ($k_F$ again denotes the Fermi momentum).  A complete expression for $F (k^2,k_F,m_F)$ can be found in Ref. \cite {KapustaT}.  We concentrate on the expression for $F (k^2,k_F,m_F)$ in the massless ($m_F=0$) limit that is a good approximation for 
ultra-relativistic fermions:
\beq
F (k^2,k_F)=  {e^2 \over 24\pi^2}  
\left ( 
16k^2_F +{k_F (4k_F^2 -3k^2)\over k}\ln ( {2 k_F  +k \over 2 k_F -k })^2
- k^2 \ln ( { k^2 -4 k_F^2\over \mu_0^2} )^2 
\right )\,.  
\label{F}
\eeq
Here $\mu_0$ stands for the normalization point that appears in the one-loop vacuum polarization diagram calculation.  The function $F$ introduces a  shift  of the pole in the propagator, corresponding to the ``electric mass'' of the photon.  This part of the pole can be incorporated via the TF approximation, as was done above.  In addition, however, the function $F$ also gives rise to branch cuts in the complex $|\k|$ plane (see \cite {Walecka} for the list of earlier references on this).
 
In analogy with the propagator found above (\ref{photonphonon}), we can decompose the static potential as follows: 
\beq
\label{photonphonon1}
\tilde V (\k, \omega=0)= \left ( \k^2 +m_\gamma^2 +F\right )^{-1} - \left ( 
\k^2 +m_\gamma^2+ F+ {\k^2 (\k^2+m_\gamma^2+F)^2\over 4M^4} \right )^{-1}\,.
\eeq
The first term in (\ref{photonphonon1}) is just the instantaneous screened-Coulomb (Yukawa) potential  of a massive photon with the one-loop polarization correction.  Our main interest is at distances smaller  than $m_\gamma^{-1}$. A sphere of radius $m_\gamma^{-1}$ encloses many particles within its  volume since $m_\gamma^{-1}\gg J_0^{-1/3}$.  At these scales, the first term in (\ref{photonphonon1}) can be approximated by:
\beq
\label{CoulF} 
{1\over \k^2 +F}\,.
\eeq
The above expression has a regular pole corresponding to the acquired ``electric'' mass of the photon  due to the polarization diagram.  The contribution of this  pole would give rise to an exponentially 
decaying potential $e^{-m_{\rm el}r}/r$, where $m_{\rm el} \sim e \mu_f$.  This is just an ordinary Debye screening.

However, as was mentioned above, the expression (\ref {CoulF}) also has branch cuts in the complex $|\k|$ plane for $k=\pm 2k_F$. These branch cuts give rise to the additional terms in the static potential which are not exponentially suppressed,  but instead have an oscillatory behavior with a power-like decaying envelope.  In a non-relativistic theory they're known as the Friedel oscillations \cite {Walecka}. In the relativistic theory they were calculated in Refs. \cite {Sivak,KapustaT} (we follow here \cite {KapustaT} and for simplicity ignore the running of the coupling constant due to the vacuum loop):
\beq
\label{oscp}
\Delta V= {Q_1Q_2 \alpha_{\rm em}^2 \over 4\pi} {{\rm sin}(2k_Fr)\over k_F^3r^4}\,.
\eeq
These branch cuts have a physical interpretation: since there is no mass gap in the fermion spectrum, a photon can produce a near-the-Fermi-surface particle-hole pair of an arbitrarily small energy and a momentum close to $\pm 2k_F$. The imaginary part of the one-loop photon polarization diagram  should include the continuum of such near-the-Fermi-surface pairs. These are reflected as logarithmic branch cuts in the expression for $F$.

Thus, if the phonon term (the second term) on the r.h.s. of (\ref{photonphonon1}) were absent one would have a power-like behavior (\ref{oscp}) of the static potential at scales $r \lesssim m_\gamma^{-1}$. The phonon term, however, significantly reduces the strength of this potential. The result for this term can be calculated by directly taking the Fourier transform of (\ref{V00}).  The dominant contribution comes from the branch cuts at $k=\pm 2k_F$.  Drawing the contours  around these cuts in the upper half plane of complex $|\k|$ \cite {Walecka,KapustaT},  one deduces the result.  In the approximation $M\gg k_F \gg m_\gamma$, which is relevant for our system, a static potential between like charges scales as 
\beq
\label{oscpo}
\Delta V \simeq  {4 Q_1Q_2 \alpha_{\rm em}^2 \over \pi} {k_F^5~ {\rm sin}(2k_Fr)\over M^8 r^4}\,.
\eeq
The potential (\ref{oscpo}) is a generalization of the Friedel potential to the case when in addition to the fermionic excitations there are also collective modes due to the  charged condensate.  As this contribution to the overall potential is a result of a subtraction between the conventional Friedel term and the long-range oscillating term due to a phonon, its magnitude is suppressed by a factor of $16(k_F/M)^8$, as compared to what it would have been in a theory without the condensed charged bosons (see \cite{Walecka} for the discussion of the conventional Friedel potential, and Ref. \cite{Dolgov2} for its recent detailed study in the presence of the charged condensate at finite temperature.)\footnote{Note that for spin-dependent interactions the same effects of the charged condensate would give a generalization of the Ruderman-Kittel-Kasuya-Yosida (RKKY) potential  \cite{RKKY}.}

Nevertheless, $\Delta V$ in (\ref{oscpo}) dominates over the exponentially suppressed part of the total potential found in (\ref{potential}), for separations between probe particles large enough for the effective field theory description to be applicable. The net static interaction in the charged condensate, set by (\ref{oscpo}), is very weak.  It is, however, still much stronger than gravitational interaction between a pair of light nuclei.  Although formally  $\Delta V $ in (\ref {oscpo}) is proportional to $\alpha_{\rm em}^2$,  to a good approximation it is  independent of  $\alpha_{\rm em}$ since $M^8 \propto g^4 (m_H J_0)^2$.

The net potential takes the form
\beq
V_{stat}= \alpha_{\rm em}  {Q_1Q_2} \left (  { \esp^{-Mr}\over \,r}
{\rm cos} (Mr)\, + {4 \alpha_{\rm em} 
\over \pi} { k_F^5{\rm sin}(2k_Fr)\over M^8r^4}\right )\,. 
\label{potential2}
\eeq
The first, exponentially suppressed term modulated by a periodic function, is due to the cancellation between the screened Coulomb potential and that of a phonon \cite{GGRR4}.  The existence of such a potential 
due to cancellation between the photon and phonon exchanges was 
first pointed out in Ref. \cite {Lee2}, in the context of 
superconductivity.

More important, however, is the second term in (\ref{potential2}) that has a long-range \cite {GGRR4}.  It dominates over the exponentially suppressed term in (\ref{potential2}) for scales of physical interest, and exhibits the  power-like behavior modulated by a periodic function. 

This potential (\ref{potential2}) is not sign-definite.  In particular, it can give rise to attraction between like charges;  this attraction is due to collective excitations of both fermionic and bosonic degrees of freedom.  This represents a generalization of the Kohn-Luttinger effect \cite{Kohn} to the case where, on top of the fermionic excitations, the collective modes of the charged condensate also contribute.

In the charged condensate Cooper pairs of electrons can also be formed.  However, the corresponding transition temperature and the magnitude of the gap, are suppressed by a factor of ${\rm exp} (-1/e_{eff}^2)$, where $e_{eff}^2$  is proportional to the value of the inter-electron potential that contains both screened Coulomb and phonon exchange.  The fact that this potential has an attractive domain, though very small, can be seen from the static potential found above (\ref{potential2}); the latter is suppressed by a power of a large scale $M$.  Furthermore, taking into account the frequency dependence of the propagator in the Eliashberg equation does not seem to change qualitatively the conclusion of a strong  suppression of the Green's function and pairing temperature.  

Hence, even though the bosonic sector (condensed nuclei) is superconducting at  reasonably high temperatures $\lesssim 10^{6}~K$, interactions with gapless fermions could dissipate the superconducting currents. Only at extremely low temperatures, exponentially close to the absolute zero, could the electrons also form a gap leading to the superconductivity of the whole system.  In the present work we consider temperatures at which electrons are not condensed into Cooper pairs, and ignore the finite temperature effects.

Finally we note that the magnetic interactions are not screened at the scale $M^{-1}$.  Instead, as is clear from the Lagrangian (\ref{LpertF}), the magnetic interactions are screened at scale of the Compton wavelength of the massive photon.  This scale will end up being much greater that the average inter-particle separation $d$. 

This may seem somewhat puzzling since the one-loop fermion correction to the transverse  part of the photon  propagator may be expected to introduce corrections that are of the order  $(m_0^2-m_\gamma^2)$, which would dominate over any effect of the order $m^2_\gamma$.
This would be the case, for instance, in a plasma where the fermion loop would determine the plasma frequency.  However, in the charged condensate the issue is more subtle,  
as was already seen in the zero-zero component of the photon propagator:  the modification of the propagator due to the condensate does not  simply reduce to a shift of the pole by $m_\gamma^2$, but instead gives rise to an additional infrared-sensitive momentum-dependent  term in the propagator. This modification is such that the fermion loop correction to the real part of the pole is  negligible in comparison with the contribution due to the charged condensate.
In other words,  the fermion-loop correction to the mass of the the longitudinal mode is negligible (the correction is of the order of $m_0^2/m_H^2 \ll 1$). The transverse mode has to have a mass equal to that of the longitudinal mode, which is determined by $m_\gamma$, since there is no difference between the transverse and longitudinal modes at zero momentum.

\chapter{Helium White Dwarf Stars}
\label{ch3}

\section{Condensation versus crystallization}
White dwarf stars represent a final evolutionary state of low mass stars.  Because white dwarfs have exhausted their thermonuclear fuel, they are no longer supported against collapse by the heat generated by fusion.  Instead they are stabilized by the degeneracy pressure of the electrons balancing against the gravitational attraction of the ions.  As a result white dwarf stars are very dense - they are roughly of the size of the Earth and their mass is on the order of a solar mass.  Their central mass densities range from $\sim (10^6-10^{9})$ g/cm$^3$, with most of them falling near the lower edge of this interval.  Their cores consist of a neutral system of electrons and nuclei (ions), the interparticle separation between nuclei being in between the atomic scale (Angstr\"om $\sim 10^{-8}$ cm) and the nuclear scale (Fermi $ \sim 10^{-13}$ cm).  Thus the electrons and the nuclei are unable to form neutral atoms, yet nuclear effects can be considered insignificant.

What, then, is the state of matter in the cores of these stars?  To a certain extent, the answer is known - it depends on the temperature $T$.  At high temperatures the equilibrium state is a plasma of negatively charged electrons and  positively charged nuclei.  However, at lower temperatures the system undergoes significant changes.  Generally, as the star cools below a critical temperature $T_{cryst}$, the plasma becomes strongly coupled enough for the ions to crystallize \cite{MestelRuderman}.  This is the case for the majority of white dwarfs whose cores are composed of carbon or oxygen nuclei.  However, for certain systems, in particular those whose cores are composed of helium nuclei, the de Broglie wavelengths of the ions begin to overlap significantly before the crystallization temperature is reached.  In this case quantum effects can prevent the crystallization transition.  Instead, the interior of the helium dwarfs may form a quantum state in which the electrons form a degenerate Fermi liquid and the charged helium-4 nuclei condense into a macroscopic state of large occupation number -- the charged condensate.  

To show this, let us first consider the cooling process in the core of a white dwarf while ignoring quantum mechanical effects for the nuclei.  We will treat the electrons as a degenerate Fermi liquid, as the Coulomb energy of a pair of electrons in a white dwarf is smaller than the Fermi energy.  For a typical dwarf star, below a certain temperature $T_{cryst}$, the Coulomb energy of a pair of ions will significantly exceed their classical thermal energy.  In order to minimize their Coulomb energy, the ions will arrange themselves into a crystal lattice \cite{MestelRuderman}.  The crystallization temperature is characterized by the dimensionless ratio of the average Coulomb energy of a pair of ions to the thermal energy (see, e.g., \cite{Shapiro}) 
\beq
\label{Gamma}
\Gamma \equiv {E_{Coulomb}\over \tfrac{2}{3}E_{Thermal}}= {(Ze)^2\over 4 \pi d } \, \frac{1}{T}\,.
\eeq
Here $e$ denotes the electric charge, $Ze$ is the charge of a nucleus, and we have set the Boltzmann constant $k_B=1$.  The interparticle separation of the nuclei is given by $d \equiv (4 \pi J_0/3Z)^{-1/3}$ where $J_0$ is the electron number density.  The second equality in (\ref{Gamma}) assumes the validity of the classical approximation.

Numerical studies have indicated that when the temperature drops low enough so that $\Gamma \gtrsim 180$, the ion plasma becomes strongly coupled enough for the system to crystallize (for earlier works see Refs. \cite{MestelRuderman,vanHorn}, for later studies see \cite{Ichi,DeWitt,Chabrier} and references therein).  This has direct relevance to white dwarf stars: it is expected that in most white dwarfs, consisting of carbon, oxygen, or heavier elements, the crystallization transition takes place in the process of cooling (for a review, see, e.g., Ref. \cite{WDs}). 

The above discussions were classical.  As the star cools, quantum effects may become significant before crystallization sets in.  For instance, in certain systems the zero-point energy of the ions can exceed the classical thermal energy $T$ before the crystallization temperature is reached.  The Debye temperature $\Theta_D$ gives the scale at which these considerations become relevant:
\beq
\label{Debye}
\Theta_D\equiv \Omega_p \, ,~~~~ \Omega_p= Ze \,\left (\frac{J_0}{Z m_H} \right )^{1/2},
\eeq
where $\Omega_p$ is the ion plasma frequency and $m_H$ is the ion mass.  The plasma frequency is related to the zero-point energy of the ions by $\omega_0 = \Omega_p/\sqrt{3}$.  Often, $\theta_D$ may  significantly exceed the crystallization temperature $T_{\rm cryst}$ \cite{Ashcroft}.  In such cases, quantum zero-point oscillations should be taken into account.  This seems to delay the formation of quantum crystal, lowering $T_{\rm cryst}$ from its classical value at most by about $\sim 10 \%$ \cite{Chabrier}.  Since this is a small change, we will ignore it in our estimates.

There is another scale at which quantum effects must be taken in consideration.  For some systems, in particular those with light ions or those that are extremely dense, the de Broglie wavelengths of the ions will begin to overlap before the crystallization temperature is reached.  The de Broglie wavelength is given by $\lambda_{\rm dB} = 2 \pi/ |\k|$ where $\k^2/2m_H = 3T/2$.  The overlap of the de Broglie wavelengths indicates that the quantum mechanical uncertainty in the position of the particles is greater than the average inter-particle separation.  The critical  temperature  $T_c$ at which this overlap occurs is given by
\beq
\label{Tc}
T_c \simeq \frac{4 \pi^2}{3 m_Hd^2}\,.
\eeq
Below $T_c$ the ions must be treated as indistinguishable particles obeying Bose-Einstein statistics.  The statistics of indistinguishable bosons at low temperatures lead to the condensation of the bosons into the lowest energy quantum state. We discussed this in detail in chapter 1.  Thus for systems in which $T_c > T_{cryst}$, instead of crystallizing, the system may condense into a macroscopic zero-momentum quantum state with large occupation number -- the charged condensate.\footnote{Following arguments of the previous chapter, we take the condensation temperature to be roughly that of the critical temperature: $T_{condens} \simeq T_c$.}

Dynamically, the condensation proceeds in the conventional manner: at temperatures higher than the condensation temperature $T_c$ most of the states are in thermal modes, and $\mu_s(T)$ is less than $m_H$.  As the temperature drops, $\mu_s(T)$ increases.  (Here we ignore the temperature dependence of $m_H$ which would be present due to the quantum loop effects).  As $T \to T_c$ a significant fraction of the modes ends up in the zero momentum ground state, while $\mu_s(T_c)$ asymptotes to $m_H$.

We wish to consider a system in which the quantum effects described above are maximal.  Because helium-4 nuclei are lighter and have lower charge than carbon and oxygen nuclei, we expect them to be more sensitive to quantum effects.  While the majority of white dwarf stars have cores composed of carbon or oxygen nuclei, helium-core white dwarfs constitute a small sub-class of dwarf stars (see, \cite {Eis,24} or references therein).  Most of helium dwarfs are believed to be formed in binary systems, where the removal of the envelope off the dwarf progenitor red giant by its binary companion happened before helium ignition, producing a remnant that evolves to a white dwarf with a helium core. Helium dwarf masses range from $\sim 0.5~M_\odot$ down to as low as $(0.18-0.19)~M_\odot$, while their envelopes are mainly composed of hydrogen.   The system is long-lived as the helium-4 nuclei are stable w.r.t. fission.  Furthermore, some nuclear reactions that could contaminate the helium-4 cores by their products are suppressed.  One of these is the neutronization process due to inverse beta-decay.  In our case the electrons are not energetic enough to reach the neutronization threshold of the helium-4 nucleus, which is about $20~{\rm MeV}$.  Moreover, we would expect that the rate of helium fusion via the triple alpha-particle reaction is suppressed as the so-called pycnonuclear reaction (i.e., nuclear fusion due to zero-point oscillations in a high density environment) rates are exponentially small \cite {Shapiro}.  Hence, the cores of these white dwarfs are expected to be dominated by helium-4 for very long time after their formation.

\begin{table}[ht]
\caption{Typical physical quantities for a helium white dwarf}
\vspace{0.1cm}
\centering
\begin{tabular}{|c| c|}
\hline\hline
Physical quantity  & Numerical value \\ [0.5ex]
\hline
Mass density     & $10^6$ g/cm$^3$  \\
Electron number density & $(0.13 ~{\rm MeV})^3$ \\
Separation between atomic nuclei & $ 10^{3}$ fm \\
Debye temperature  & $4\cdot 10^6$ K \\
Critical temperature  & $ 10^{6}$ K \\
Crystallization temperature   & $3 \cdot 10^5$ K \\  [1ex]
\hline
\end{tabular}
\end{table}
 
 Table 3.1 summarizes the relevant physical quantities for a helium white dwarf star with a typical mass density $\rho \simeq 10^6$ g/cm$^3$.   We see that the condensation temperature $10^6$ K is significantly greater than the crystallization temperature $3 \cdot 10^5$ K.  Thus we expect that the system of the helium-4 nuclei and electrons may not solidify, but may condense instead.  For helium-core white dwarfs with even denser cores, the effect is greater still.

\section{The non-relativistic EFT}
As we discussed in chapter 1, for a large number of quanta that form a macroscopic state, such as the helium nuclei in the charged condensate, the state can be described in terms of an effective field theory of the order parameter, and its long wavelength fluctuations.  The effective field theory should be constructed based on the fundamental symmetries of the physical system and the properties of the interactions involved.  The relativistic effective Lagrangian adopted in the previous chapter (\ref{lagr}) is not the most appropriate description for the charged condensate in the cores of helium white dwarf stars, although captures many of its significant features.  It is overly restrictive in that it enforces Lorentz invariance - a symmetry we do not expect the low energy system of helium-4 nuclei and electrons to preserve.  In addition it contains a heavy mode of mass which we would expect to be beyond the scope of a low energy theory.  In this section we discuss an low energy, non-relativistic effective Lagrangian description of charged condensation, and in particular its application to the system of helium-4 nuclei and electrons described above. We will see that in this formalism the charged condensate retains the distinctive features found in the previous chapter, namely equivalent Lorentz-violating dispersion relations for the massive photon and the same strong screening of electric charge.

We focus on the zero-temperature limit, even though realistic temperatures in helium white dwarfs are well above zero.  The validity of the zero-temperature approximation is justified {\it a posteriori} as follows: the spin-0 nuclei undergo condensation to the zero-momentum state; while they do so they cannot excite their own phonons since the latter are gapped with the magnitude of the gap being greater than the condensation temperature.  On the other hand, the condensing charged bosons can and will excite thermal fluctuations in the fermionic sector that is gapless.  Therefore, all the thermal fluctuations will end up being stored in the fermionic quasiparticles near the Fermi surface.  For the latter, however, the finite temperature effects aren't significant since their Fermi energy is so much higher than the temperature,  $T/J_0^{1/3} \ll 10^{-2}$.  We note that the finite temperature effects, in a general setup with  condensed bosons, were calculated in Refs. \cite {Dolgov,Dolgov2}.

The electron/nuclei system in the cores of the white dwarf has three relevant mass scales: the mass of a nucleus $m_H$, the electron mass $m_e$ and the electron chemical potential $\mu_F$.  The mass of the nuclei is significantly greater than the other two mass scales.  For the effective Lagrangian construction we consider scales that are well below the heavy mass scale $m_H$, but somewhat above the scale set by $m_e$ and $\mu_F$.  Hence the electrons are described by the Dirac Lagrangian, while for the description of the nuclei we will use a charged scalar order parameter $\Phi(x)$.

The effective Lagrangian for $\Phi$ must satisfy the following requirements: (i) it must be consistent with the symmetries of the physical system: translational, rotational, and Galilean symmetries, a global $U(1)$ symmetry for the conserved scalar number, and a local gauge invariance;  (ii) it should reproduce the standard Schr\"odinger equation for the order parameter in lowest order in the fields;  (iii) it should obey an algebraic relation between the conserved current density and the momentum density:  $J_j = (e/m_H) T_{0j}$;  (iv) it should give an appropriate spectrum of Nambu-Goldstone bosons in the decoupling limit.  Such a Lagrangian was first proposed by Greiter, Wilczek and Witten (GWW) \cite {GWW} in a context of superconductivity:
\beq
\label{Lgww}
{\cal L}_{eff} = {\cal P} \left ({i\over 2} ( \Phi^*  D_0 \Phi -  (D_0 \Phi)^* \Phi)-{| D_j  \Phi|^2  \over 2m_H} \right )\,,
\eeq
where $D_0 \equiv (\partial_0  - i2e A_0)$,  $ D_j \equiv ( \partial_j - i 2e A_j) $.  The function ${\cal P}(x)$ stands for a general polynomial function of its argument.  The coefficients of this polynomial are dimensionful numbers that are inversely proportional to powers of a short-distance cutoff of the effective  field theory, ${\cal P}(x) = \sum^{\infty}_{n=0} C_n x^n $. 

The Yukawa terms are forbidden by symmetries in this case. However, in general one should also add to the Lagrangian terms $\mu_{NR} \Phi^*\Phi$, $\lambda (\Phi^*\Phi)^2/m_H^2$, \linebreak $\lambda_1 (\Phi^*\Phi){\bar \psi}\psi/(m_H J_0^{1/3})$, and other  higher dimensional operators that are consistent with all the symmetries and conditions that lead to (\ref {Lgww}) (the Yukawa term is not). Here $\mu_{NR}$ denotes a non-relativistic chemical potential for the scalars.  These terms are not important for the low-temperature spectrum of small perturbations we're interested in, as long as  $\lambda,\lambda_1 \lesssim 1$ and  $J_0\ll m^3_H$.  However, near  the phase transition point it is the sign of $\mu_{NR}$ that would  distinguish between the broken and symmetric phases, so these terms should be included for the discussion of the symmetry restoration.  We also note that the scalar part of  (\ref {Lgww}) is somewhat similar to the Ginzburg-Landau (GL) Lagrangian for superconductivity. However, there are significant differences between them, one such difference being that the coherence length in the GL theory is many orders of magnitude greater than the average interelectron separation, while in the present case, the ``size of the scalar''  $\Phi$ is smaller that the average interparticle distance.

The GWW effective Lagrangian can also describe charged condensation, given the appropriate rescaling and reinterpretation of the parameters of the theory.  In the condensate where the VEV of $\Phi$ is nonzero, we can express $\Phi$ in term of a modulus and phase: $\Phi = \Sigma\, {\rm exp}(i\Gamma)$.  Written in terms of fields $\Sigma$ and $\Gamma$, the effective Lagrangian (\ref{Lgww}) takes the following form: 
\beq
\label{Lgww1}
{\cal L}_{eff} = {\cal P} \left ((2e A_0-\partial_0 \Gamma) \Sigma^2 - \frac{1}{2 m_H} (\nabla_j  \Sigma )^2 -\frac{1}{2 m_H}(2e A_j-\partial_j \Gamma)^2 \Sigma^2  \right )\,.
\eeq
For the fermions we adopt  the usual Dirac Lagrangian with a relativistic chemical potential
\beq
\label{lagrF}
{\cal L}_F = {\bar \psi}(i\gamma^\mu D_\mu -m_F)\psi + \mu_F \psi^\dagger \psi\,, 
\eeq
where the covariant derivative for the electrons is defined  as $D_\mu = \partial_\mu +ie A_\mu $.  The electron chemical potential $\mu_F$ is equal to the Fermi energy at zero temperature $\mu_F=\epsilon_F= [(3 \pi^2 J_0)^{2/3}+m_F^2]^{1/2}$.  There exists a static solution that follows from Lagrangians (\ref{Lgww1}) and (\ref{lagrF}).  In the unitary gauge $\Gamma = 0$, this solution has the form
\beq 
\label{Lgwwsol}
2e\Sigma^2 =e J_0\,,~~~ A_\mu=0 \, ,~~~{\cal P}^{\prime}(0)=1\,.
\eeq 
Since on the solution the argument of (\ref{Lgww1}) is zero, the condition ${\cal P}^{\prime}(0)=1$ is satisfied by any polynomial function for which the first coefficient is normalized to one: ${\cal P} (x) = x + C_2 x^2+...$.  The above solution describes a neutral system in which the helium-4 charge density $2e\Sigma^2$ exactly cancels the electron charge density $-eJ_0$.  

\vspace{0.3cm}

Let us now turn to the issue of fluctuations about the classical solution.  We express $\Sigma$ in terms of a perturbation $\tau$ above the condensate value:
\beq
\label{Sigmapert}
\Sigma(x) = \sqrt{m_H} \left(\sqrt{\frac{J_0}{2 m_H}}+\tau(x) \right) \, .
\eeq
The Lagrangian, including the gauge field kinetic term, expanded to second order in fields becomes:
\beq
\label{Lnrpert}
{\cal L}_2 &=& -{1 \over 4} F_{\mu\nu}^2-{1 \over 2} (\partial_j \tau)^2+{1 \over 2}\left(C_2 m_H J_0 m_\gamma^2 +\frac{e^2}{\pi^2} (3\pi^2 J_0)^{2/3}\right) A_0^2 \\  \nonumber
&& - {1\over 2} m_\gamma^2 A_j^2+2m_H m_\gamma A_0 \tau \, .
\eeq
Here we have included fluctuations in the electron number density via the TF approximation.  This gives rise to the second mass-like term, i.e. the Debye mass, in front of $A_0^2$.  The photon mass is defined as $m_\gamma \equiv 2e \, \sqrt{J_0/2m_H}$.

We can compare this Lagrangian to the one obtained in the relativistic theory (\ref{LpertF}).  (Here the perturbation $b_\mu$ is equal to $A_\mu$, as for the gauge field we are expanding above the background $A_\mu = 0$.)  The first difference between the two Lagrangians is the lack of a time derivative for $\tau$ in the non-relativistic case.  This is because the heavy field is decoupled in the low energy limit. This, obviously, 
does not change the conclusions about the static potential discussed 
in the previous chapter.

The second difference between (\ref{Lnrpert}) and (\ref{LpertF}) is the presence of the expansion coefficient $C_2$ in the former.  This is due to the fact that the non-relativistic theory for the scalar sector is not required to be Lorentz invariant.  While the mass term due to the electron fluctuations also breaks the degeneracy between the ``electric" and ``magnetic" masses of the gauge field, this is because the electron number density sets  a preferred Lorentz frame.  In the absence of this term, a Lorentz invariant Lagrangian for the scalars would have to have $C_2 = 1/(m_H J_0)$ and thus the ``electric" and ``magnetic" masses would be equal.  Within the non-relativistic effective Lagrangian approach, however, the value of $C_2$ cannot be fixed {\it a priori}.  It must be fixed instead by the physics of the particular system being described.

As long as $C_2$ is chosen so that the ``electric" mass of the photon does not exceed the scale $M$, then our results will not greatly depend on the particular choice of $C_2$.  To calculate the spectrum of perturbations for the non-relativistic theory, we set by hand $C_2 = 1/(m_H J_0)$ for simplicity.  Then the Lagrangian for small perturbations above the condensate in the non-relativistic theory is identical to the Lagrangian for small perturbations in the relativistic theory up to a time derivative for $\tau$:
\beq
\label{Lnrpert2}
{\cal L}_2 = -{1 \over 4} F_{\mu\nu}^2-{1 \over 2} (\partial_j \tau)^2+{1 \over 2} m_0^2 A_0^2 - {1\over 2} m_\gamma^2 A_j^2+2m_H m_\gamma A_0 \tau \, ,
\eeq
where again ${m}_0^2 \equiv  m^2_\gamma+ {e^2/ \pi^2 \, (3\pi^2 J_0)^{2/3} }$.  Calculating the spectrum of perturbations from this Lagrangian gives two transverse components of the gauge field which propagate according to the usual massive dispersion relation $\omega^2 = m^2_\gamma+\k^2 $.  In addition there is a longitudinal mode with the dispersion relation:
\beq
\label{omeganr}
\omega^2 = m_\gamma^2+ \frac{m_\gamma^2 \k^4}{m_0^2 \k^2+4 M^4} \simeq m_\gamma^2+ \frac{\k^4}{4 m_H} \, .
\eeq
For the last equality we have taken the low momentum limit.  This dispersion relation should be compared to those found for the relativistic theory in the $m_H \gg m_\gamma$ limit (\ref{pmdisp}).  We find that the heavy mode found  in (\ref{pmdisp}) is no longer present in the low energy theory, as we would expect.  In addition, the term proportional to $-m^2_\gamma \k^2/4m_H^2$ that appears in the longitudinal dispersion relation for $\omega_-$ in (\ref{pmdisp}) does not appear in the non-relativistic theory, since it is of a higher order in $1/m_H$. It was this term that 
was responsible for the roton-like behavior of the small fluctuations discussed in the previous section.

Nevertheless, the longitudinal excitation exhibits the same mass gap as seen in the relativistic theory and thus the bosonic part of the system still satisfies the Landau criterion for superfluidity.  Moreover, the phonon dispersion relation retains the same essential form that gives rise to the unusual potential for a static probe charge placed in the condensate (\ref{potential}).  Thus the screening of electric charge in the non-relativistic theory is identical to that found in the previous chapter.

\section{White dwarf cooling}
The presence of charged condensation in the cores of some white dwarf stars could have significant observational consequences. In this subsection 
we follow Ref. \cite {GGDP}. Consider the phonon contribution to the specific heat of a white dwarf in which the core has crystallized.  A conventional phonon in a crystal has a dispersion relation $\omega \propto | \k |$ at low $|\k|$.  As a result, the contribution of the phonon to the specific heat scales as $T^3$.  In contrast, in the condensate, the phonon dispersion relation exhibits a mass gap $m_\gamma$ as discussed above.  As a result, contributions of the bosons to the specific heat at low temperatures are exponentially suppressed as ${\rm exp} (-m_\gamma/T)$.  Since for typical densities in helium white dwarfs we have $m_\gamma \simeq 3$ KeV, the suppression at temperatures below $10^6$ K will be $\sim {\rm exp}(-30)$.  

The electrons exhibit largely the same behavior in both the crystallized and condensate phase.  At the relevant temperatures they form a degenerate Fermi gas with gapless excitations near the Fermi surface.  As a result their contribution to the specific heat scales linearly with temperature.  Thus, in the case of the crystallized cores, the contribution of the electrons to the specific heat is subdominant compared to the contribution coming from the massless crystal phonon.  In the condensate, however, the gapless excitations of the electrons will be the dominant contribution to the specific heat.  This difference in the specific heat between the crystallized phase and the condensate phase has a significant impact on the cooling of helium white dwarf stars.

Using the approach of \cite{Shapiro}, and following \cite {GGDP} we consider an over-simplified model of a reference helium star of mass $M=0.5~M_\odot$ with the atmospheric mass fractions of hydrogen, and heavy elements (metallicity) respectively equal to
\beq
 X\simeq 0.99, \quad  \quad Z_m \simeq (0.0002-0.002)~.
\label{eq01}
\eeq
The lower value of the metallicity $ Z_m \simeq 0.0002$ is appropriate for the recently discovered 24 He WDs in NGC 6397 \cite{24}, but for completeness, we consider a wider range for this parameter. 

It is straightforward to find the following expression for the cooling time of a star in the classical regime \cite {Shapiro}
\beq
t_{He}=\frac{k_B}{CAm_u}\left [ \frac{3}{5}(T_f^{-\frac{5}{2}}-
T_0^{-\frac{5}{2}})+Z\frac{\pi^2}{3}\frac{k_B}{E_F}
(T_f^{-\frac{3}{2}}-T_0^{-\frac{3}{2}})\right ],
\label{eq03}
\eeq
where $T_f$ and $T_0$ denote the final and initial core temperatures. The first term in the brackets on the right hand side corresponds to cooling due to the classical gas of the ions and the second term corresponds to the contribution coming from the Fermi sea. The latter is sub-dominant in the range of final temperatures that we are interested in (the factor Z in front of this term is due to $Z$ electrons per ion). Since  $T_ f \ll T_0$, the age of a dwarf star typically doesn't depend on the initial temperature.  Neglecting the fermion contribution, we find the time that is needed to cool down to the critical temperature $T_f=T_c$
\beq
t_{He}=\frac{3}{5}\frac{k_BT_c M}{Am_uL(T_c)}\simeq (0.76 - 7.6)~\text{Gyr}\, ,
\label{ages}
\eeq
where an order of magnitude interval in (\ref {ages}) is due to the interval in the envelope metallicity 
composition given in (\ref {eq01}).  We also find the corresponding luminosities
\beq
L(T_c) \simeq(10^8~erg/s)\frac{M}{M_\odot}
\left ( \frac{T_c}{\text{K}}\right )^{{7/2} }\simeq 1.5\cdot 
(10^{-4}-10^{-5}) L_{\odot}\,,
\eeq
which are in the range of observable luminosities ($L_{\odot}\simeq 3.84\cdot 10^{33} ~erg/s$).

After condensation, the specific heat of the system dramatically drops as the collective excitations of the condensed nuclei become massive and  ``get extinct''.  The contribution from the Fermi sea, which is strongly suppressed by the value of the Fermi energy, becomes the dominant one.   The phase transition itself will take some time to complete, and the drop-off in specific heat will not be instantaneous.

\begin{figure}[h]
\begin{center}
\epsfxsize=6in
\epsffile{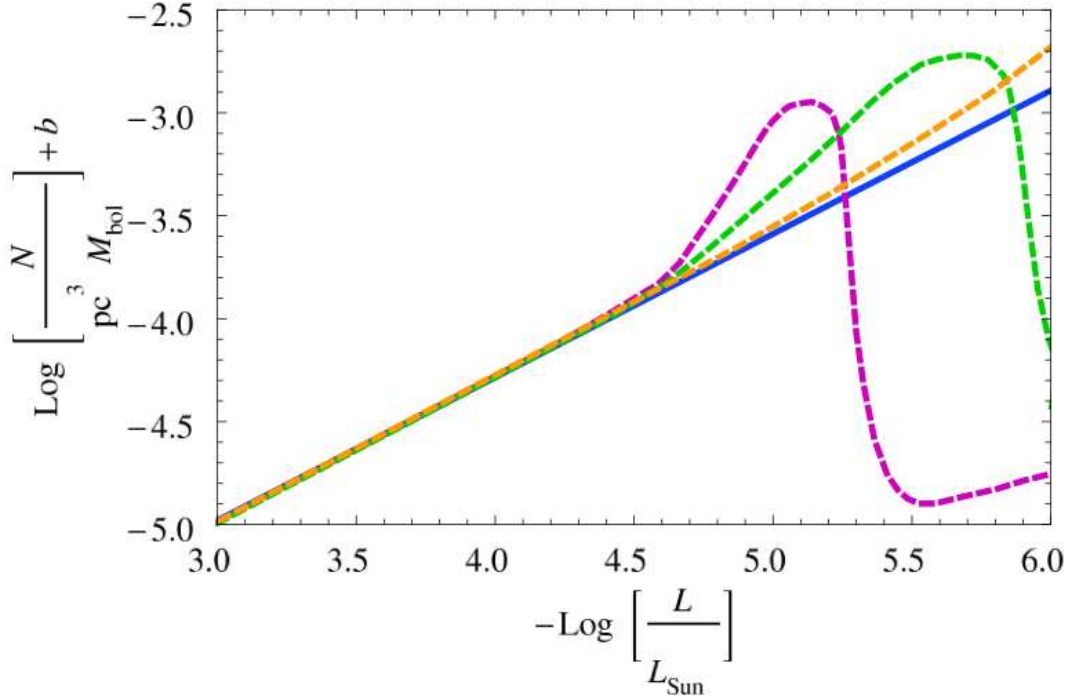}
\end{center}
\caption{A schematic sketch of the luminosity function for helium-core white dwarfs, taken from Ref. \cite {GGDP}. The absolute normalization of the function is set by the constant $b$ which is determined by their formation rate. The blue line represents the Mestel regime. The shape of the luminosity function near the condensation point depends on the details of the corresponding phase transition.}
\label{lum}
\end{figure}

In the zeroth approximation, we can regard the transition to be very fast, and retain only the fermion 
contribution to the specific heat below $T_c$. Then, the expression for the age of the star for $T_f<T_c$, reads as follows
\beq
t_{He}'=\frac{k_B}{CAm_u}\left [ \frac{3}{5}(T_c^{-\frac{5}{2}}-
T_0^{-\frac{5}{2}})+Z\frac{\pi^2}{3}\frac{k_B}{E_F}
(T_f^{-\frac{3}{2}}-T_0^{-\frac{3}{2}})\right ].
\label{eq02}
\eeq
Notice the difference of (\ref {eq02}) from  (\ref {eq03}) -- in the former $T_f <T_c$ and it is $T_f$ that enters as final temperature in the  fermionic part, while  $T_c$ should be taken as the final temperature in the bosonic part. 

From the ratio of ages, $\eta= {t_{He}/ t_{He}'}$, for two identical helium dwarf stars, with and without the 
interior condensation, we deduce that the charged condensation substantially increases the rate of  cooling-- the age could be twenty times less than it would have been without the condensation phase \cite {GGDP}.

The condensation of the core would induce significant deviations from the classical curve for helium white dwarfs. What is independent of the uncertainties involved in these discussions, is the fact that the luminosity function (LF) will experience a significant drop-off after the charged condensation phase transition is complete. This is due to the ``extinction''  of the bosonic quasiparticles below the phase 
transition point. In fact, the LF  will drop by a factor of $\sim 200$. This is illustrated schematically in Fig. \ref{lum}, taken from Ref. \cite{GGDP}. 

This may be relevant for the results of Ref. \cite {24}. 
The latter work reported the observed termination of a sequence of the 24 He WD's found in a nearby  globular cluster NGC 6397.  
The termination takes place  before the limiting luminosity 
is reached. Moreover, it takes place before termination is seen in the 
carbon- and/or oxygen-core white dwarfs observed in the same cluster. 
The direct astrophysical explanation -- the disruption of the helium-core WD containing  binaries by collisions  --
does not seem to be favored for the local densities in the 
environment  where the sequence was observed \cite {24}. It is tempting to speculate that the observed termination of the sequence may be 
a signature of fast cooling due to the charged condensation. 
For these discussions see Ref. \cite {GGDP}.

\chapter{Vortex Structure}
\label{ch4}

\section{The vortex solution}
In this chapter we argue that the charged condensate has properties somewhat similar to type II superconductors. In particular, we show that it can admit solutions that are similar to the Abrikosov  vortices \cite{Abri}, originally found in the Ginzburg-Landau model of superconductivity, and later recovered in the relativistic abelian Higgs model in \cite{NO}.  The vortex solution is a topologically stable configuration, characterized by a nonzero winding number of the phase of the complex scalar field.  Asymptotically, the scalar field is given by $\phi \sim v \esp^{i \theta}$, where $v$ is the vacuum expectation value of the field and $\theta$ is the azimuthal coordinate.  Like the Abrikosov vortex, the charged condensate vortex carries a quantized magnetic flux.  The vortex solution has a higher energy density than the pure condensate solution.  However, in the presence of a sufficiently high external magnetic field, it becomes energetically favorable for the charged condensate to form vortices.

The obtained vortex line solution exhibits the following structure: it has a narrow cylindrical core where the scalar field changes significantly from a zero to a nonzero value; this core is surrounded by a broad halo in which the magnetic flux is confined. The width of the latter region is determined by the penetration depth (i.e., the photon Compton wavelength).  We refer to the system of the core and the halo as  the flux-tube, or the vortex line.  This structure is similar to that of the Abrikosov solution.  However, unlike the latter, our solution also carries a  profile of the electrostatic potential within the halo, while this potential is exponentially small outside of the halo (i.e., the flux tube is charge neutral).  Hence, in terms of differential equations, one has to solve three coupled equations instead of the two required in the Abrikosov \cite{Abri}, or Nielsen-Olesen cases \cite{NO}.

If charged condensation exists in the cores of some white dwarf stars, it is possible that vortex structure exists as well.  Surface magnetic fields have been found in white dwarf stars ranging from $10^3$ to $10^9$ Gauss.  We will show that the presence of such strong magnetic fields should be sufficient for the the creation of vortices in the condensed cores of white dwarfs.  The existence of vortices in the cores of these stars could have observational consequences.

The organization of this chapter is as follows:  In this section we consider a generic system of charged scalars and oppositely charged fermions in the context of the relativistic formalism presented in chapter 2.  We fix the phase of the scalar field to be of the vortex-type and solve the corresponding equations of motion.  We compare our solutions to those found in the abelian Higgs model.  In section 4.2 we consider corrections to our solutions due to the dynamics of the fermions.  In section 4.3 we consider the effects of an external magnetic field on the charged condensate and determine the magnitude of the external field for which it becomes energetically favorable to form vortices.  In section 4.4 we treat specifically the case of helium-4 nuclei and electrons.  We describe the system in the context of the non-relativistic low energy effective field theory developed in the previous chapter.  We discuss the applicability of the vortex solutions found in this section (4.1) to the helium-4 nuclei and electron system.  We also consider the effect of a constant rotation on the condensate of helium-4 nuclei.

\vspace{0.3cm}

To find the vortex solution we use the relativistic formalism developed in chapter 2.  We assign a charge of $+2e$ to the scalars and $-e$ to the fermions in anticipation of the helium-4 nuclei and electron system to be discussed later in section 4.4.  However, for now we keep our considerations general.  The scalar field considered below could be any fundamental scalar field, or a composite order parameter suited for 
a problem at hand. Our conclusions are independent of the specific charge assignment.   The relativistic Lagrangian is given by (see \ref{lagr})
\beq
\label{lagrV}
{\cal{L}}'=-\tfrac{1}{4} F_{\mu\nu}^2 + 
\tfrac{1}{2}(\partial_{\mu}\sigma)^2+
\tfrac{1}{2}(2eA_\mu+\mu_s \delta_{\mu 0}-\partial_\mu \alpha)^2 \, \sigma^2- 
\tfrac{1}{2} m_H^2 \, \sigma^2 - e A^{\mu} J_{\mu} \, .
\eeq
The equations of motion that follow from (\ref{lagrV}) are:
\beq
\label{eom1V}
-\partial^\mu F_{\mu \nu}& =& 2e(2eA_\nu+\mu_s \delta_{\nu 0}-\partial_\nu \alpha) \sigma^2-eJ_\nu \, ,  \\
\label{eom2V}
\Box \, \sigma &= &[(2eA_\mu+\mu_s \delta_{\mu 0}-\partial_\mu \alpha)^2-m_H^2] \, \sigma \, .
\eeq
In the unitary gauge the charged condensate solution is given by
\beq
\label{staticsol}
\langle 2 e A_0 \rangle+\mu_s= m_H,  ~~~~\langle \sigma \rangle = \sqrt{\frac{J_0}{2m_H}} \, ,
\eeq
and the photon mass is $m_\gamma = 2e \, \sqrt{J_0/2m_H}$.  In the following we will consider a system whose net charge is zero and thus we will set $\langle A_0 \rangle= 0$ and $\mu_s = m_H$.

To find the charged condensate solution (\ref{staticsol}) in the unitary gauge, we fixed the phase of the scalar field to zero.  We now consider a configuration where the phase is not set to zero, nor can it be set to zero everywhere by a non-singular gauge transformation.  The requirement that the scalar field be single-valued everywhere is satisfied by demanding that the change in phase around a closed loop be an integer multiple of $2 \pi$.  In a system with cylindrical symmetry, this is satisfied by setting $\alpha = n \theta$, where $\theta$ is the azimuthal coordinate and $n$ is an integer.  This phase can be removed everywhere by a gauge transformation $A_\mu \rightarrow A_\mu +\partial_\mu (n \theta)$, except at the origin where the gauge transformation would be singular.  The solutions of the equations of motion (\ref{eom1V}), (\ref{eom2V}) where the phase is fixed to $\alpha = n \theta$ are vortex-type solutions.

At the origin $r=0$, in order for the scalar field $\phi$ to be well-defined we must have $\sigma = 0$.  Here $r$ is the 2D radial coordinate.  Far from the origin however, we expect the solutions to recover the condensate values (\ref{staticsol}).  At large $r$ then, the gauge field takes the form $2eA_j \rightarrow \partial_j \alpha$, or equivalently $A_\theta \rightarrow n/(2er)$.  From this form of the vector potential, it follows that this configuration has a quantized magnetic flux ${\it \Phi}$ that is related to the integral of $A_j$ around a closed loop at infinity:
\beq
\label{flux}
{\it \Phi} = \oint {\bf{A \cdot}} \d {\bf{l}}= \oint A_\theta r \d \theta  = \frac{2 \pi n}{2e}\, .
\eeq
The magnetic flux is quantized in units of $n$.  The quantization of flux implies the stability of the vortex configuration, although it may be possible for a high $n$ vortex to decay into multiple vortices of smaller $n$.

To solve the equations of motion (\ref{eom1V}), (\ref{eom2V}) for the vortex configuration we switch notation to dimensionless variables.  The resulting equations are governed by a single parameter $\kappa$, the ratio of the mass of the scalar to the mass of the photon in the condensate: $\kappa = m_H/m_\gamma$.  This parameter $\kappa$ is the equivalent to the Ginzburg-Landau parameter in the theory of superconductors which gives the ratio of the penetration depth to the coherence length.  For the helium white dwarf star, if we take the mass of the helium-4 nuclei to be roughly $m_H =3.7$ GeV and the electron density to be $J_0 \sim (0.15-0.5 {\rm{~Mev}})^3$, then we have $\kappa \sim 10^6$.  In our derivations below we frequently take the large $\kappa$ limit.

We define $x \equiv m_\gamma r$, set $A_r = A_z = 0$ and $\mu_s = m_H$, and perform the following change of variables:
\beq
\label{changevar}
m_\gamma A(x) & \equiv& 2exA_\theta(x) \, , \\
m_\gamma F(x) & \equiv& 2e \sigma(x) \, , \\
m_H B(x)& \equiv& \mu_s+2eA_0(x)\, .
\eeq
In terms of these new variables equations of motion (\ref{eom1V}), (\ref{eom2V}) become
\beq
\label{eomA}
x \, \frac{d}{dx}\left(\frac{1}{x}\frac{dA}{dx}\right)&=&F^2 (A-1) \, , \\
\label{eomF}
-\frac{1}{x}\frac{d}{dx}\left(x\frac{dF}{dx}\right)&=&\left[\kappa^2(B^2-1)-\frac{n^2}{x^2}(A-1)^2\right]F \, , \\
\label{eomB}
\frac{1}{x}\frac{d}{dx}\left(x\frac{dB}{dx}\right)&=&F^2 B -1 \, .
\eeq
The boundary conditions are set by requiring that the solutions asymptote to the condensate solutions for large $r$, while for $r=0$ we have $A_\theta = \sigma = dA_0/dr = 0$:
\beq
\label{bc}
\begin{array}{llll}
{\rm{For }} ~x \rightarrow 0: & A(x) \rightarrow 0, &  F(x) \rightarrow 0, & \frac{dB}{dx}  \rightarrow 0 \, . \\
{\rm{For }} ~x \rightarrow \infty: & A(x) \rightarrow 1, & F(x) \rightarrow 1, & B(x) \rightarrow 1\, . \\
\end{array}
\eeq

\vspace{0.3cm}

We can compare these expressions to those obtained in the usual abelian Higgs model.  Suppose that instead of Lagrangian (\ref{lagr}) we had the abelian Higgs Lagrangian:
\beq
\label{LAH}
{\cal L}^{\rm AH} =-\frac{1}{4} F_{\mu\nu}^2 + 
\frac{1}{2}(\partial_{\mu}\sigma)^2+
\frac{1}{2}(2eA_\mu-\partial_\mu \alpha)^2 \, \sigma^2- 
\frac{\lambda}{4} (\sigma^2 - v^2)^2 \, .
\eeq
Using the same change of variables as above and defining $m_H^{\rm AH} = \sqrt{\lambda} v$, $m_\gamma^{\rm AH} = 2ev$, the equations of motion are:
\beq
\label{eomAAH}
x \, \frac{d}{dx}\left(\frac{1}{x}\frac{dA}{dx}\right)&=&F^2 (A-1) \, , \\
\label{eomFAH}
-\frac{1}{x}\frac{d}{dx}\left(x\frac{dF}{dx}\right)&=&\left[\kappa^2(1-F^2)-\frac{n^2}{x^2}(A-1)^2\right]F \, .
\eeq
The equation of motion for the vector potential, expressed via $A$, is the same as in the charged condensate model.  In the equation for the scalar field, the $\sigma^4$ term in the abelian Higgs model gets replaced in the charged condensate model by a term that depends on the electric potential.  In addition, in the charged condensate equations the electric potential is generally not zero and not constant and has its own equation to satisfy.

\vspace{0.3cm}

Let us first examine the asymptotic behavior of the solutions to the condensate equations for $x \rightarrow \infty$.  Far from the origin we expect the fields to be very close to their condensate values.  Then, on the r.h.s. of equation (\ref{eomA}), it follows that $A(x)-1 \equiv a(x)$ is very small.  If we consider this equation only to first order in small fields then we can approximate the scalar field on the r.h.s. of (\ref{eomA}) as $F \simeq 1$.  The solution for $A$ that obeys the appropriate boundary conditions is
\beq 
\label{Asol}
A(x)=1-c_a x K_1(x) \, .
\eeq  
Here $c_a$ is a constant to be determined by the matching of the solutions and $K_1(x)$ is the modified Bessel function of the second kind.  In the large $x$ limit this solution for $A$ becomes
\beq
\label{Alarge}
A(x) \rightarrow 1-c_a \sqrt{\frac{\pi x}{2}} {\rm{e}}^{-x} \,.
\eeq

To find the asymptotic behavior of $B$ and $F$ we expand these fields in terms of perturbations above the condensate values, $B(x)=1+b(x)$ and $F(x)=1+f(x)$, and we assume that $b(x), f(x) \ll1$.  We then substitute these expressions as well as expression (\ref{Alarge}) into the equations for $B$ and $F$ and keep only terms linear in the perturbations $b(x)$ and $f(x)$.  These two equations can be combined to obtain a fourth order differential equation for $b(x)$.  Using the ansatz $b(x) = c_b \, x^s \, {\rm{e}}^{-kx}$ where $c_b$, $s$, and $k$ are as yet undetermined constants, we can find the particular and homogenous solutions for $b(x)$ in the large $x$ limit.  We also take $c_a \simeq 1$ which we will justify later.  For the particular solution we find
\beq
\label{bp}
b_p(x) = \frac{\pi n^2}{4 (\kappa^2+3)}\frac{{\rm{e}}^{-2x}}{x} \, .
\eeq
For the homogenous solution we find $s=-1/2$ and $k^2 = (1\pm \sqrt{1-16 \kappa^2})/2$.  In the limit that $\kappa$ is very large, the solution becomes
\beq
\label{bh}
b_h(x) = \frac{{\rm{e}}^{-\sqrt{\kappa}x}}{\sqrt{x}} \left[c_1 \sin(\sqrt{\kappa}x) +c_2 \cos(\sqrt{\kappa}x)\right] \, ,
\eeq
for some constants $c_1$ and $c_2$.  The complete solution is then
\beq
\label{Btot}
B(x) = 1 + \frac{\pi n^2}{4(\kappa^2+3)}\frac{{\rm{e}}^{-2x}}{x} + \frac{{\rm{e}}^{-\sqrt{\kappa}x}}{\sqrt{x}} \left[c_1 \sin(\sqrt{\kappa}x) +c_2 \cos(\sqrt{\kappa}x)\right]\, .
\eeq
The solution for $F(x)$ can be found once $B(x)$ is known:
\beq
\label{Ftot}
F(x) = 1 + \frac{3\pi n^2}{8(\kappa^2+3)}\frac{{\rm{e}}^{-2x}}{x} - \frac{\kappa {\rm{e}}^{-\sqrt{\kappa}x}}{\sqrt{x}} \left[c_1 \cos(\sqrt{\kappa}x) -c_2 \sin(\sqrt{\kappa}x)\right]\, .
\eeq
Here $c_1$ and $c_2$ are the same integration constants that appear in the expression for $B(x)$.  

As $\kappa$ is large, the second term in the above expressions for $B$ and $F$ dominates the asymptotic behavior.  For $x\rightarrow \infty $ we have
\beq
\label{Basym}
B(x) \rightarrow 1 + \frac{\pi n^2}{4(\kappa^2+3)}\frac{{\rm{e}}^{-2x}}{x} \, , \\
\label{Fasym}
F(x) \rightarrow 1 + \frac{3\pi n^2}{8(\kappa^2+3)}\frac{{\rm{e}}^{-2x}}{x} \, .
\eeq
The asymptotic behavior for the vector potential and the scalar field are similar to that for the abelian Higgs model:
\beq
\label{AFasymHiggs}
A^{\rm AH}(x)  = 1-c_a \sqrt{\frac{\pi x}{2}} {\rm{e}}^{-x} \, , ~~~~~~
F^{\rm AH}(x) = 1 - \frac{c_a^2 \pi n^2}{4(\kappa^2-2)}\frac{{\rm{e}}^{-2x}}{x} +c_f \frac{\esp^{-\sqrt{2}\kappa x}}{\sqrt{x}}\, . 
\eeq
The vector potential $A$, and thus the magnetic field, are the same in both the charged condensate and abelian Higgs models.  The asymptotic behavior of the scalar field in the abelian Higgs model in the large $\kappa$ limit is dominated by the ${\rm{e}}^{-2x}$ term, as in the charged condensate model.

Notably, this is not the asymptotic behavior for the Abrikosov vortex given in the Nielsen-Olsen paper \cite{NO}.  This discrepancy was first pointed out by  L. Perivolaropoulos in \cite{Peri}.  The incorrect asymptotic behavior is obtained if one similarly expands $A(x)$ as $A(x) = 1+a(x)$ and only keeps terms linear in $a(x)$.  This is because the last term in (\ref{eomF}) and the last term in (\ref{eomFAH}) are quadratic in $a(x)$ and yet, due to the different exponential dependence of the perturbations $a(x)$, $b(x)$ and $f(x)$, these terms can be dominant over terms which are linear in $b(x)$and $f(x)$.  Linearizing $A$ gives the correct asymptotic behavior of the fields only in the limit that $\kappa$ is small.

The second term in the full expressions for $B$ and $F$ and thus the asymptotic behavior of both fields is due strictly to the presence of a nonzero magnetic field.  In the absence of any magnetic field, the screening of any small perturbation of the fields above their condensate values vanishes as $\esp^{-\sqrt{\kappa}x}=\esp^{-\sqrt{m_H m_\gamma}r}$ (as was found in \ref{b0sol}) and (\ref{potential})).  

\vspace{0.3cm}

The above asymptotic expressions are valid as long as $x \gg 1$, or equivalently $r \gg 1/m_\gamma$.  We consider now the intermediate region $1/\sqrt{\kappa} \ll x \ll 1$, or equivalently $ 1/M \ll r \ll 1/m_\gamma$ where again we have defined $M \equiv \sqrt{m_H m_\gamma}$.  At distances much larger than $1/M$ we assume that the scalar field $F$ is still close to its condensate value.  Thus expression (\ref{Asol}) is still valid for $A$.  In this regime then $n^2/x^2 (A-1)^2 \simeq n^2/x^2$.  The equations for $B$ and $F$ become
\beq
\label{eomBF2}
\frac{1}{x}\frac{d}{dx}\left[x\frac{dB}{dx}\right]=F^2 B-1\, , ~~~~~
 \frac{1}{\kappa^2}\left[\frac{1}{x}\frac{d}{dx}\left(x\frac{dF}{dx}\right)-\frac{n^2}{x^2}F\right]=(1-B^2)F\, .
\eeq
The solutions are straightforward to find:
\beq
\label{BFsol}
B(x)=\left(1+\frac{n^2}{\kappa^2 x^2}\right)^{1/2}\,,  ~~~~
F(x)=\left(1-\frac{n^2}{2\kappa^2 x^2}+\frac{2n^2}{\kappa^2 x^4}\right)^{1/2} \, .
\eeq
Alternatively, we can once again expand $B$ and $F$ above their condensate values, $B(x)=1+b(x)$ and $F(x)=1+f(x)$, and solve for $b(x)$ and $f(x)$.   The homogenous solutions for $b(x)$ and $f(x)$ are the same as those given above with the same coefficients $c_1$ and $c_2$.  Solving for the particular solutions gives the full solutions in the linearized approximation:
\beq
\label{Blin}
B(x)& =& 1 + \frac{n^2}{2\kappa^2 x^2}+ \frac{{\rm{e}}^{-\sqrt{\kappa}x}}{\sqrt{x}} \left[c_1 \sin(\sqrt{\kappa}x) +c_2 \cos(\sqrt{\kappa}x)\right]\, , \\
\label{Flin}
F(x)& =& 1 - \frac{n^2}{4 \kappa^2x^2}+ \frac{n^2}{\kappa^2x^4}- \frac{\kappa {\rm{e}}^{-\sqrt{\kappa}x}}{\sqrt{x}} \left[c_1 \cos(\sqrt{\kappa}x) -c_2 \sin(\sqrt{\kappa}x)\right]\, .
\eeq
The coefficients $c_1$ and $c_2$  are needed to perform the matching.  However, as we'll see below,  these coefficients will turn out not to be exponentially large, and hence
solutions (\ref {Blin}) and (\ref {Flin}) approximate well the solutions  in (\ref {BFsol}).

The approximations made to find both the homogenous and particular solutions break down as $x$ approaches $1/\sqrt{\kappa}$.  Moreover, $f(x)$ becomes of order 1 at $x \sim 1/\sqrt{\kappa}$ and thus the linear approximation in general no longer holds below this scale.

\vspace{0.3cm}

Finally, we'd like to  solve in the $r \rightarrow 0$ limit.   Before we do so, however, 
we emphasize that validity of this procedure needs some justification.  
The interparticle separation is given by $d \propto J_0^{-1/3}$.  
This corresponds to $x \propto 1/\kappa^{1/3}$.  
At distances shorter than this $x$ we expect that an effective field theory 
would break down and thus it would make little physical sense to solve 
the equations (\ref{eomA}), (\ref{eomF}) and (\ref{eomB}) in this regime.  
Moreover, the scale $1/M$ is typically shorter than  the interparticle separation 
$d$, hence, particles at these scales cannot in general be modeled by  
a smooth distribution. 

Both fermions and bosons are in a condensate state in which the location of individual 
particles has uncertainties much greater than  the interparticle separation. Hence the 
latter notion loses its meaning as a microscopic characteristic of the system.
For this, we'll  still approximate particle distributions by smooth functions
all the way down to the scale $\sim 1/M$, which  is a dynamically determined short-distance 
scale at which weakly coupled expansion breaks down \cite {GGRR1}.  As to solving at scales less that $1/M$, we regard this as a purely mathematical exercise aimed at finding the matching of the asymptotic solutions for the corresponding differential equations for all values of the coordinate $x$. 

Taking $A$, $B$, and $F$ to be series expansions in small $x$ obeying the appropriate boundary conditions, the solutions to (\ref{eomA}), (\ref{eomF}) and (\ref{eomB}) are 
\beq
\label{A0}
A(x) &=& a_0 x^2-\frac{f_0^2}{8} x^4 \, , \\
\label{B0}
B(x) &=& b_0 -\frac{x^2}{4} \, , \\
\label{F0}
F(x) &=& f_0 \left[x-\frac{1}{8}\left(\kappa^2(b_0^2-1)+2 a_0\right)x^3\right] \, .
\eeq 
For simplicity we have solved for the case that the winding number $n=1$.\footnote{For $n \neq 1$, the leading term in the expansion for $F$ will be $\propto x^n$.  The expression for $B(x)$ remains unchanged and the leading term in the expansion for $A(x) = a_0 x^2$ is also the unchanged.}  The coefficients $a_0$, $f_0$, and $b_0$, as well as the coefficients $c_a$, $c_1$, and $c_2$ can be determined by matching the above solutions to those in the intermediate region, given by (\ref{Asol}), (\ref{Blin}) and (\ref{Flin}).

To determine the physically appropriate matching radius, we first use Gauss's law to 
find the charge of the vortex solution.  The number density of fermions in the 
vortex $J_0$ is constant and is the same as the number density of fermions 
in the normal condensate phase.  We have fixed it so by hand, but will justify this later.  
The scalar number density is given by $\tfrac{1}{2} J_0 BF^2$ and varies as a function of $x$.  
Therefore it is not in general equal to its condensate value $\tfrac{1}{2} J_0$.  
The variation of the scalar number density away from its condensate value can lead to a 
net charge density of the vortex core within the vortex halo.  
In particular, there are two competing effects.  In the intermediate 
region $1/\sqrt{\kappa} \ll x \ll 1$, both $B$ and $F$ are above their condensate 
values, thus the scalar number density is greater than the scalar number density in 
the condensate.  As $x \rightarrow 0$, however, $F \rightarrow 0$ and the scalar number 
density drops to zero, significantly below the condensate value.  The matching radius 
should be chosen so that these two effects combine to give the appropriate 
charge density as determined by Gauss's law.  

From Gauss's law we can calculate  the average charge density of the vortex inside 
radius $x=1$.  As is usually the case, we can determine the net charge enclosed in a region 
knowing only the form of the potential at the boundary of that region.  Equation (\ref{BFsol}) 
gives the potential in the intermediate region independent of matching coefficients $c_1$ and 
$c_2$.  This form of the potential, together with Gauss's law, allows us to calculate the net 
charge  of the vortex at $x=1$ independent of the matching conditions and the $x \rightarrow 0$ 
solutions.

Gauss's law is given by equation (\ref{eom1V}):
\beq
\label{Gauss}
\nabla^2 A_0 = 2e (2eA_0+\mu_s-\dot{\alpha})\sigma^2- e J_0 \, .
\eeq
The r.h.s. of the above equation is the charge density.  Integrating both sides of the above expression over the volume of the vortex and dividing by the total volume gives the average charge density inside distance $x$:
\beq
\label{Qenc}
\frac{Q_{\rm{enc}}}{V} = 2eJ_0\frac{1}{x}\frac{dB}{dx} \, .
\eeq
Here $V$ is the volume equal to the length of the vortex times the cross-sectional area and the r.h.s. is evaluated at the boundary of the vortex.  Using expression (\ref{BFsol}) for $B(x)$ at $x=1$, the average charge density inside $x=1$ is $Q_{\rm{enc}}/V=- 2e J_0/\kappa^2$.  The negative sign indicates a dearth of scalars in this region, but, as $\kappa$ is very large, this is a small correction to the overall average charge density of scalars $\propto e J_0$.  To check this result one can likewise use the asymptotic solution for $B$, expression (\ref{Btot}), at $x=1$.  Assuming that the coefficients $c_1$ and $c_2$ are not exponentially large and thus these terms are not dominant in the solution for $B$ at $x =1$, one finds $Q_{\rm{enc}}/V \propto- 2e J_0/\kappa^2$.  This is consistent with the previous result.  Farther out, $B(x)-1$ is exponentially suppressed thus the net charge of the vortex approaches zero as $x$ becomes large.

We can now use this result to determine the matching radius $R$.  Given the smallness of the average charge density found inside $x = 1$, the excess of scalars in the intermediate region of the vortex must cancel the shortage of scalars in the $x \rightarrow 0$ region to great accuracy.  Using expressions (\ref{BFsol}) for $B$ and $F$, it can be shown that this happens when $R \simeq 1/\sqrt{\kappa}$.  Thus we use this as our matching radius $R$ in what follows.

We start by matching the solution for $A(x)$ in (\ref{A0}) and its first derivative with its solution in the intermediate region (\ref{Asol}).  Taking the matching radius to be small, $R \ll 1$, gives:
\beq
\label{aco}
a_0 = - \frac{1}{2} \left[\gamma +\ln\left(\frac{R}{2}\right) \right] \, , ~~~~ c_a =1+\frac{R^2}{4} \, ,
\eeq
where $\gamma$ is the Euler-Mascheroni constant.  As long as $R$ is less than one, $a_0$ is positive.  Moreover, we see that we were justified in taking $c_a \simeq 1$ in our previous calculations.  The magnetic field is given by 
\beq
\label{mag}
H = \frac{m_\gamma^2}{2e} \, \frac{1}{x} \frac{dA}{dx} \, .
\eeq
Near the origin the magnetic field is of order $m_\gamma^2/(2e)$.   For $x > 1/\sqrt{\kappa}$ it is given by $m_\gamma^2 K_0(x)/(2e)$, where $K_0(x)$ is the modified Bessel function.  For $x  \gg 1$, i.e. for $r  \gg 1/m_\gamma$, the magnetic field is exponentially small.

To find the remaining coefficients, we use $a_0$ found above and match (\ref{B0}) and (\ref{F0}) and their first derivatives to the appropriate solutions in the intermediate region (\ref{Blin}), (\ref{Flin}).  We now take the matching radius to be $R=1/\sqrt{\kappa}$.  The solution with the lowest energy is one in which the scalar field $F(x)$ is identically zero in the region $x < R$.  The corresponding coefficients are
\beq
\label{bfco}
\begin{array}{cclccl}
\vspace{0.2cm} 
b_0 &=&1+\frac{7}{4 \kappa}  \, ,&~~~~c_1& =&\kappa^{-5/4}\, \esp \left(2 \cos(1)+\sin(1)\right) \, , \\ 
f_0 &=&0\, , &~~~~c_2&=&\kappa^{-5/4}\, \esp \left(\cos(1)-2\sin(1)\right) \, .
\end{array}
\eeq
In Fig.\ref{fig} below, the fields are plotted for small $r$ and for $\kappa \sim 10^6$.  The radius $r=1/M$ corresponds to the matching radius $x=R=1/\sqrt{\kappa}$.  The radius $r = d$ denotes the interparticle separation $d = J_0^{-1/3}$.  Unlike the magnetic field, the potential and scalar field approach their condensate values for $x > 1/\sqrt{\kappa}$.  This is in contrast to the abelian Higgs model in which the scalar field is close to its condensate value for $x > 1/\kappa$, i.e. for $r > 1/m_H$.

\begin{figure}[h!]
\begin{center}
\subfigure[Scalar field as a function of radius]{\epsfig{figure=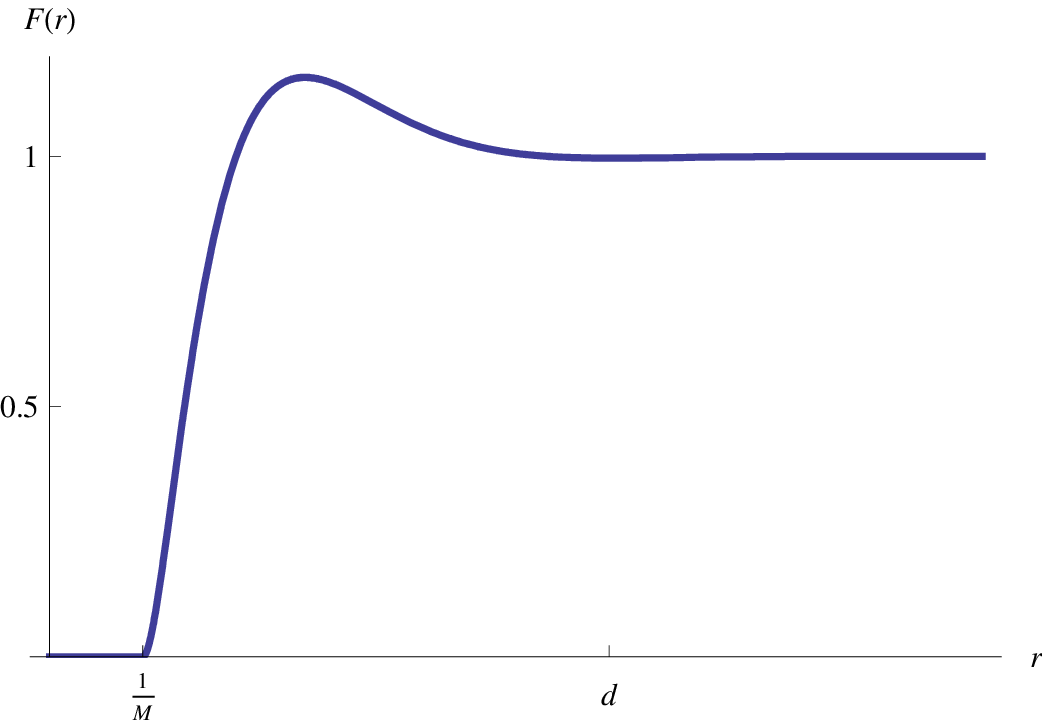,width=.45\textwidth}}
\hskip .15in
\subfigure[Potential as a function or radius]{\epsfig{figure=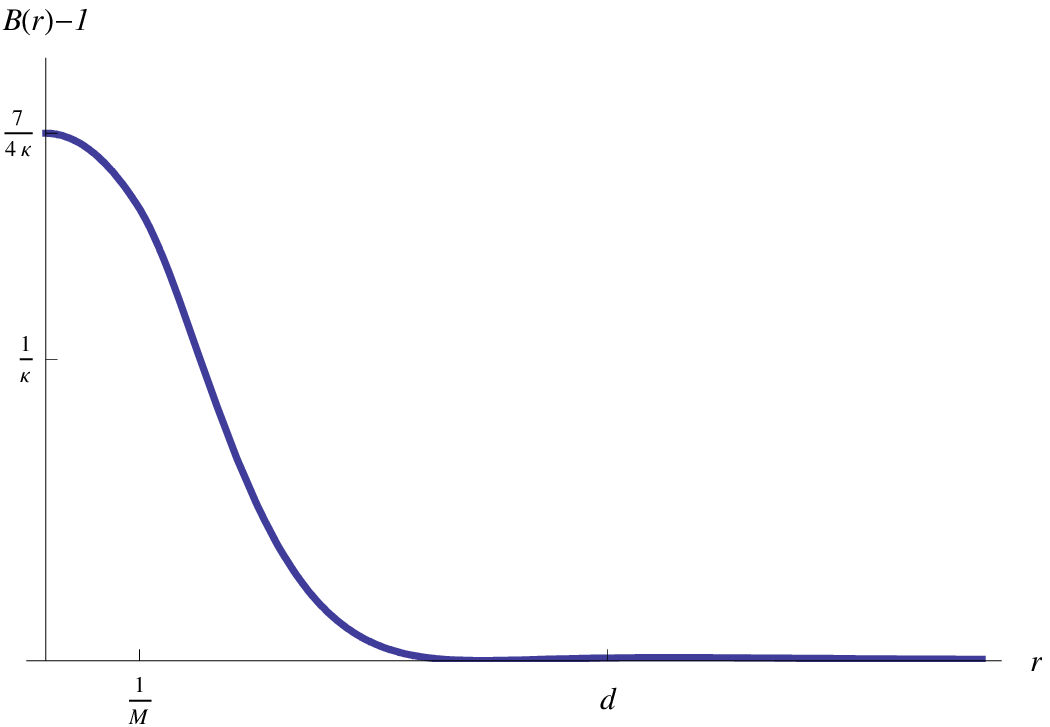,width=.45\textwidth}}\\
\end{center}
\caption{Small $r$ solutions for the scalar field and electric potential.}
\label{fig}
\end{figure}

\section{Fermion dynamics}
In our discussions above we have treated the fermions as a fixed charge background $J_\mu  = J_0 \delta_{\mu 0}$.  We relax this assumption now and introduce dynamics for the fermions via the Thomas-Fermi approximation.  The fermion dynamics are governed by the constant chemical potential $\mu_F$ given in (\ref{muF}).  The local number density of fermions is determined by the Fermi momentum: $J_0(x) = p_F^3(x)/(3 \pi^2)$.  In this way the number density of the fermions $J_0$ gets related to the electric potential $A_0$.  For relativistic fermions
\beq
\label{J0V}
J_0(x) = \frac{1}{3 \pi^2} (\mu_F-e A_0(x))^3 \, .
\eeq
The chemical potential gets fixed by the value of the fermion number density in the condensate phase, where $\langle A_0 \rangle = 0$.  If $\bar{J}_0$ represents the number density of fermions in the condensate, then $\mu_F = (3 \pi^2 \bar{J}_0)^{1/3}$.  The photon mass $m_\gamma$ is also defined in terms of $\bar{J}_0$: $m_\gamma \equiv 2e \sqrt{\bar{J_0}/2m_H}$.  In the vortex phase $J_0(x) \rightarrow \bar{J}_0$ for large $x$.

To include the effects of an $x$-dependent $J_0$ into our equations, (\ref{J0}) gets incorporated into the equations of motion (\ref{eom1}).  As a result the equation of motion for $B(x)$ (\ref{eomB}) gets modified.  In the linearized equations, the effect is the addition of a new term for $b(x)$ with a coefficient which scales as $\propto m_H/\mu_F$.  However, it turns out that this new term does not contribute significantly to the solutions.  This is because, in the fourth order differential equation for $b(x)$, terms with coefficient $m_H/\mu_F \propto \kappa^{2/3}$ are subdominant compared to terms with coefficient $\kappa^2$.  Accordingly, the solutions found above in the intermediate and asymptotically large regions are still valid.  It can be shown that the $x \rightarrow 0$ solutions (\ref{A0}), (\ref{B0}), (\ref{F0}) are also unaffected.  In physical terms, the inclusion of the fermion dynamics via the Thomas-Fermi approximation gives rise to ordinary Debye screening.  This screening is subdominant compared to other screening effects in the condensate (see \cite{GGRR3}).  Moreover, the profile of $A_0(x)$ away from the core and within the halo is very shallow, giving rise to a very mild dependence of the charge density on $x$. Hence, the latter can be approximated by a constant, as was done in the previous sections.

As mention in section 2.3, the TF approximation does not capture the possibility of exciting gapless modes near the Fermi surface.  To include this effect we calculated the one-loop correction to the gauge boson propagator, which gives corrections to the static potential $A_0$.  We are interested in how this correction compares to the potential found in the intermediate region of the vortex (\ref{Blin}).  To estimate its magnitude, we consider a toy model of the vortex.  We find the potential due to a wire of constant linear charge density $\lambda_0$ located at $r=0$.  The linear charge density of this wire is set by the characteristic charge of the vortex: since the scalar charge density varies significantly from its condensate value $e J_0$ at scales $r < 1/M$, it follows that at short distances the linear charge density of the vortex can be approximated by
\beq
\label{lambda0}
\lambda_0 = \frac{e \pi J_0}{M^2} \, .
\eeq
At large distances the vortex is effectively neutral, as mention above.  Thus we expect the one-loop contribution to the static potential to be irrelevant at large scales.

In three dimensions, the charge density of the source is given by
\beq
\label{J0source}
J_0^{\rm source} (r,\theta,z) = \frac{\lambda_0}{\pi r} \delta(r) \, .
\eeq
The static potential is determined from this source and from the $\{00\}$ component of the gauge boson propagator $D_{00}$:
\beq
\label{convol}
A_0(r) = \int d^3r' \, D_{00}(r-r') J_0^{\rm source}(r') \, .
\eeq
The one-loop correction to the propagator due to this was found in (\ref{oscpo}):
\beq
\label{D00r}
D_{00}(\bar{r}) = \frac{\alpha_{\rm em}}{\pi^2} \frac{k_F^5 \sin(2k_F\bar{r})}{M^8 \bar{r}^4} \, .
\eeq
Here $\bar{r}$ represents the 3D radius in spherical coordinates, as opposed to the 2D radius $r$.

Using this expression together with expression (\ref{J0source}) in equation (\ref{convol}), the correction to the static potential is
\beq
\label{A0lamdba}
A^{\prime}_0(r) = \frac{\alpha_{\rm em}}{\pi^2} \,\frac{ \lambda_0 k_F^5}{M^8} \,
 \int_{-\infty}^\infty dz' \, \frac{\sin(2k_F\sqrt{z'^2+r^2})}{(z'^2+r^2)^2} \, .
\eeq
An upper bound on the potential can be found by taking $\sin(2k_F\sqrt{z'^2+r^2}) \rightarrow 1$.  After integrating, this gives
\beq
\label{A0lambdaless}
A^{\prime}_0(r) \, < \, \frac{\alpha_{\rm em}}{\pi^2} \,\frac{ \lambda_0 k_F^5}{M^8} \, \frac{\pi}{2 r^3}\, \propto \, \frac{\pi^2}{e^2} \sqrt{\frac{k_F}{m_H}} \frac{1}{m_H^2 r^3} \, .
\eeq

On the vortex solution, the leading term in the potential in the intermediate region is given by expression (\ref{Blin}):
\beq
\label{A0int}
A_0(r) = \frac{m_H}{2e} (B(r)-1) \simeq  \frac{1}{4e\,m_H r^2} \, .
\eeq
Given that both $k_F/m_H \ll1$ and $1/(m_Hr) \ll 1$, we see that the one-loop correction to the potential (\ref{A0lambdaless}) is greatly suppressed compared to the potential found in the vortex solution.  Thus the excitations of the fermions do not significantly alter the vortex solutions.

One further effect that we take into consideration is the Landau quantization of the fermion energy levels due to the presence of the magnetic field in the interior of the vortex.  In the presence of an applied magnetic field, the separation between energy levels is given by  $\omega = eH/m_F$ where $H$ is the magnetic field and $m_F$ is the fermion mass.  Near the core of the vortex where $H \simeq m_\gamma^2/(2e)$ the separation of levels of the fermions is $\omega \simeq m_\gamma^2/m_F$.  Since the photon mass $m_\gamma$ is generally much smaller than the fermion mass, this shift in energy is negligible compared to the typical energy of the fermions.

\section{Energetics and external fields}
We now turn to the question of when it is energetically favorable to form a vortex in the charged condensate.  We start by comparing the average energy density of the vortex to the energy density of the pure condensate.  Above, using Gauss's law, we found that inside the distance $x=1$ the vortex has a small negative charge density implying that in this region the average scalar number density is lower than in the condensate phase.  At distances $x \gg 1$ this charge density is exponentially suppressed indicating that the net charge of the vortex is zero and thus the total average scalar number density is the same in both the vortex phase and the condensate phase.  In calculating the average energy density of the vortex inside the distance $x =1$, we are not interested in the contribution to the energy due to the discrepancy in the number of scalars between the vortex phase and the condensate.  This contribution to the overall difference in energy vanishes at large distances.  Thus we calculate the energy density of the system using ${\cal{H}}'  = {\cal{H}} - \mu_s J_0^{\rm{scalar}}$.  The additional term effectively subtracts off the energy density due to the scalar number density.  We compare ${\cal{H}}'$ in the vortex phase to ${\cal{H}}'$ in the condensate.

The Hamiltonian density ${\cal{H}}'$ can be calculated from the Lagrangian ${\cal L}'$ (\ref{lagrV}):
\beq
\label{ham}
{\cal{H}}' =  \tfrac{1}{2}H^2 + \tfrac{1}{2}E^2 +\tfrac{1}{2}(2eA_0+\mu_s-\dot{\alpha})^2 \sigma^2-\mu_s(2eA_0+\mu_s-\dot{\alpha})\sigma^2  \, ,
\eeq
We have simplified the Hamiltonian using the equations of motion (\ref{eom1V}), (\ref{eom2V}) and have taken boundary terms to be negligible.  The magnetic field $H$ and the electric field $E$ are defined as usual
\beq
\label{em}
H= \frac{1}{r} \frac{d}{dr} (r A_\theta)\, , ~~~~E = -\frac{dA_0}{dr} \, .
\eeq
The fourth term in the Hamiltonian is exactly $- \mu_s J_0^{\rm{scalar}}$ as we would expect.  The third term is due to the energy of the scalar field.  Unlike in the abelian Higgs model, the energy density of the scalar field in the center of the vortex, i.e. in the ``normal" phase, is lower than in the condensate phase.  However, the energy density of the vortex is still greater than that of the condensate alone, due to the gradients of the scalar field and due to the intermediate region $1/\sqrt{\kappa} \ll x \ll 1$ in which the values of both the potential and the scalar field are greater than their condensate values.  This contribution to the energy density is roughly equal in magnitude to the contributions coming from the electric and magnetic fields.  On the condensate solution, the Hamiltonian density is identically zero: ${\cal{H}}'_{\rm CC} = 0$.

For large $x$, deviations away from the condensate are exponentially suppressed and thus differences in energy between the two phases are negligible.  So to find the average energy density within the vortex, we integrate the Hamiltonian density over an area of radius $x=1$ and then divide by the total area.  The average energy density within the radius $x \le 1$ is
\beq
\label{eps}
\epsilon_{\rm{ave}} = \int^1_0 2x\,dx\,{\cal{H}}' \, .
\eeq
The Hamiltonian density can be further simplified using the equations of motion (see Ref.\cite{SJTS} for more details).  In terms of the dimensionless variables defined above equation (\ref{eps}) becomes:
\beq
\label{eps2}
\epsilon_{\rm{ave}} =\frac{1}{2} m_H J_0  \int^1_0 x\,dx\,(B(x)-1) \, (3-F(x)^2) \, .
\eeq
For the region $x < 1/\sqrt{\kappa}$ we use solutions (\ref{B0}) and (\ref{F0}) and in the intermediate region $1/\sqrt{\kappa} < x <1 $ we use solutions (\ref{Blin}) and (\ref{Flin}) with the coefficients found from matching (\ref{bfco}).  Upon integration, the average energy density is
\beq
\label{eps3}
\epsilon _{\rm{ave}}= \frac{m_H J_0}{4 \kappa^2} \left(\log{\kappa}+ 14 \right) \,.
\eeq
The numerical coefficients should not be taken too literally given the approximations made in obtaining the solutions which yield the above result.  However, the overall scaling of the energy density $\epsilon _{\rm{ave}} \propto m_H J_0 (\log{\kappa})/\kappa^2$ is remarkably independent of the matching radius and other details of the solutions.  As the energy density of the condensate is effectively zero (${\cal{H}}'_{\rm CC} = 0$), the above expression represents the difference in energy between the two phases.  

To see when it is energetically favorable for the condensate to form vortices, we now consider placing the condensate in an external field $H_{\rm ext}$ pointed along the $z$-axis.  We shall see that the magnetic properties of the charged condensate resemble those of a superconductor.  In particular, when $\kappa \gg 1$, the charged condensate resembles a type II superconductor.  When an external magnetic field $H_{\rm ext}$ is applied to the condensate, below a critical value $H_{c1}$ surface currents oppose the penetration of the field and the induction $B_{\rm ind}$ is zero in the bulk of the condensate.  For $H_{\rm ext} > H_{c1}$ magnetic flux penetrates the condensate in the form of vortices.  At another critical value of the magnetic field $H_{c2}$ the normal phase is restored and the induction $B_{\rm ind}$ is equal to the applied field $H_{\rm ext}$.  In what follows we determine the critical values of the fields $H_{c1}$ and $H_{c2}$.

Given the energy density $\epsilon$ of the vortex phase above the pure condensate phase, we can find the value of the magnetic field $H_{c1}$ at which it becomes energetically favorable to form vortices.  In the absence of an external field, it is never energetically favorable to form vortices as the energy density of a vortex is greater than that of the pure condensate.  In the presence of a small external magnetic field, below $H_{c1}$, the condensate must expel the magnetic field entirely from its bulk in order to remain in the condensate phase.  This requires energy; the energy per volume needed to expel the external field is $\tfrac{1}{2} H^2_{\rm ext}$.  If vortices form in the condensate then the energy required to expel the magnetic field is smaller than if the field were to be completely expelled.  More specifically, if the vortices give rise to an average magnetic field in the condensate $B_{\rm ind}$, then the energy needed to expel the remaining magnetic field would be $\tfrac{1}{2} (H_{\rm ext}-B_{\rm ind})^2$.  The energy gained by forming vortices is the difference between this energy and the energy required to expel the magnetic field entirely.  Assuming that $B_{\rm ind}$ is small compared to $H_{\rm ext}$ near the transition point, this difference can be approximated by $\tfrac{1}{2} (2 B_{\rm ind} H_{\rm ext})$.  Thus for a high enough external field, the energy $\epsilon$ lost in creating a vortex is compensated by the energy gained in expelling a smaller magnetic field $B_{\rm ind} H_{\rm ext}$.  In order for formation to be energetically possible, we must have $\epsilon \leq B_{\rm ind} H_{\rm ext}$.  The equality determines the critical external field $H_{c1}$.  (See Ref.\cite{Abri}.)

Suppose the number of vortices per area in the condensate is given by $N$.  Then the energy density due to the formation of vortices is given by $\epsilon = N \lambda$ where $\lambda$ is the energy density per unit length of a vortex.  Using $\epsilon _{\rm{ave}}$ found above (\ref{eps2}) as the energy density of a single vortex,
\beq
\label{lambda}
\lambda = \frac{\pi}{m_\gamma^2} \epsilon _{\rm{ave}} \, .
\eeq
The induction $B_{\rm ind}$ is given by
\beq
\label{Bind}
B_{\rm ind} = N \, \oint {\bf{A \cdot}} \d {\bf{l}} = \frac{2 \pi N}{2 e} \, .
\eeq
Combining these expressions, the critical field $H_{c1} = N \lambda/(B_{\rm ind})$ is given by
\beq
\label{Hc1}
H_{c1} = \frac{m_\gamma^2}{8e} \left(\log(\kappa)+ 14 \right) \, .
\eeq
The final expression for $H_{c1}$ is independent of the number density of vortices $N$.  It follows that if it is energetically favorable to create one vortex, then it will be even more energetically favorable to create many, up to the point than interactions between vortices become significant.  At distances greater than $r = 1/m_\gamma$ we expect fields outside the vortices to be exponentially suppressed and thus the vortices to be effectively non-interacting.  So at the transition point $H_{c1}$, it is likely that the number density of vortices is of the order $N \simeq m_\gamma^2/\pi$.

If we take $J_0 \simeq (0.15-0.5 {\rm ~MeV})^3$, a reasonable value for white dwarfs, 
this gives a magnetic field of roughly $H_{c1} \simeq (10^{7}- 10^{9})$ Gauss.  
Thus, the vortex lines should be expected to  be present in the bulk 
of the helium-core white dwarf stars  with strong enough magnetic fields.
The condensed-core WD's with fields of strength less than 
$\sim 10^7$ Gauss would expel their magnetic fields. 
Such stars  are expected to constitute a significant fraction of 
the helium-core WD's.   

A sufficiently high magnetic field will disrupt the condensate entirely.  One way to approximate the magnetic field at which this transition occurs is to consider the density of vortices in high external magnetic fields.  When the cores of the vortices begin to overlap, then the scalars are mostly returned to the normal phase.  We define the core of the vortex to correspond to $x=1/\sqrt{\kappa}$, or equivalently, $r=1/M$ as this is the region in which the VEV of the scalar field drops to zero.  As $N$ is the number of vortices per area, at the transition point $N \simeq M^2/\pi$.  We define the critical external field at this point to be $H_{c2}$.  When the condensate enters the normal phase, the induction $B_{\rm ind}$ is equal to the external magnetic field $H_{c2}= B_{\rm ind} = 2 \pi N/(2e)$.  From these two expressions we find $H_{c2}$:
\beq
\label{Hc2}
H_{c2} = \frac{M^2}{e} \, .
\eeq
For $J_0 \simeq (0.15 - 0.5 {\rm ~MeV})^3$, $H_{c2} \simeq (10^{13}-10^{15})$ Gauss.  
This is well above the values of the fields expected to be present in a majority 
of white dwarf stars.  Thus the external magnetic field is unlikely to be 
large enough to push the condensate into the normal phase.

It should be noted that both $H_{c1}$ and $H_{c2}$ given above were determined at zero temperature.  Generally, we expect these expressions (\ref{Hc1}), (\ref{Hc2}) to be valid at temperatures well below the condensation temperature.  

Finally, in type II superconductors the dependence of the critical temperature 
on the magnetic field is well-approximated by $T^{\prime 2}_c/T_c^2 \simeq (H_c-H)/H_c$,
where $T^{\prime }_c$ is the transition temperature  when the magnetic field 
$H$ is present. We expect a similar relation to be valid in our case too. 
Hence, as long as the value of the magnetic field is not too 
close to either critical value,  the change of the transition temperature due 
to the magnetic field should be  small.  Near the critical values, however, the change 
of the phase transition temperatures  (from the normal to the vortex phase and from 
the vortex phase to the phase with no magnetic field) could change significantly.
The would be crystallization temperature will also change, and the charged condensation 
may or may not be favorable for close-to-critical magnetic fields.

\section{Comments on white dwarf stars}
To appropriately describe vortices in the cores of helium white dwarf stars, we should use the low energy effective Lagrangian description of the charged condensate developed in the previous chapter:
\beq
\label{LgwwV}
{\cal L}_{eff} = {\cal P} \left ((2e A_0-\partial_0 \Gamma) \Sigma^2 - \frac{1}{2 m_H} (\nabla_j  \Sigma )^2 -\frac{1}{2 m_H}(2e A_j-\partial_j \Gamma)^2 \Sigma^2  \right )\,.
\eeq
In what follows we consider how this alternative formalism changes the vortex solutions found above.

It was found in section 3.2 that, to linear order, the equations of motion for small perturbations above the condensate were the same in the low energy effective theory as in the relativistic effective theory, up to a time derivative for the scalar perturbation $\tau$.  As we are concerned with static solutions, this time derivative will not change the vortex solutions in the low energy theory.  In order to find vortex solutions in the intermediate region $1/\sqrt{\kappa} \ll x \ll 1$ and in the asymptotic region $x \gg 1$, we treated the electric potential $A_0$ and the scalar field $\sigma$ in the linear approximation, but kept higher order terms for the vector potential.  Thus to determine the applicability of the solutions found above to the non-relativistic effective theory, we should consider higher order terms in $A_j$ than the ones given in (\ref{Lnrpert}).  We also restore the phase $\Gamma$.  The equations of motion to next-to-leading-order that follow from (\ref{LgwwV}) are
\begin{align}
\label{eomeffB}
-\partial^{\mu} F_{\mu 0} &= 2e\left[1+2 C_2 \Sigma^2 (2eA_0-\partial_0 \Gamma)\right]\Sigma^2 -e J_0 \, , \\
\label{eomeffA}
-\partial^{\mu} F_{\mu j}& = 2e (2eA_j-\partial_j \Gamma) \Sigma^2 \, , \\
\label{eomeffF}
-\nabla^2 \Sigma &= \left[2 m_H (2eA_0-\partial_0 \Gamma) + 4 C_2 m_H \Sigma^2(2eA_0-\partial_0 \Gamma)^2 \right] \Sigma-(2eA_j)^2 \Sigma \, .
\end{align}
If we take $\Sigma = \sqrt{m_H} \sigma$ and $C_2 =1/(m_H J_0)$, then the first two equations of motion above (\ref{eomeffB}), (\ref{eomeffA}) are the same as in the non-relativistic case (\ref{eom1V}), up to second order in small fields.  The third equation (\ref{eomeffF}) has an extra factor of $(2eA_0)^2 \Sigma$ compared to equation (\ref{eom2V}).  However, since this term is second order in $A_0$ and we treated $A_0$ in the linear approximation, this does not alter our solutions in the intermediate and asymptotic regions.  Thus for $C_2 =1/(m_H J_0)$, the vortex solutions found above for $x \gg 1/\sqrt{\kappa}$ are also solutions for the non-relativistic effective theory.

As in the relativistic case, the solutions formally break down near $x=1/\sqrt{\kappa}$ when the change in the scalar field becomes of order $1$ and thus the linear approximation is no longer valid.  More realistically though, we do not expect the effective field theory to hold at distances shorter than the interparticle separation $x \propto 1/\kappa^{1/3}$.  The effective field theory will cease to be a valid description of the physics before reaching $x=1/\sqrt{\kappa}$.  Moreover, we do not in general expect that the low energy effective theory will obey the Lorentz invariant condition $C_2 =1/(m_H J_0)$.  Instead, $C_2$ must be fixed by the particular physics of the system.

We can use the non-relativistic formalism to consider the effects of the rotation of a white dwarf star on the helium-4 nuclei and electron system.  In this formalism the scalar number density and current density are given respectively by:
\beq
\label{scalcur}
J_0 ^{\rm{scalar}} = \Phi^* \Phi \, , ~~~~~~ J_j ^{\rm{scalar}} = \frac{-i}{2m_H}\left[(D_j\Phi)^* \Phi-\Phi^*(D_j \Phi) \right] \, .
\eeq 
The number density is related to the current density by $J_j ^{\rm{scalar}} = J_0 ^{\rm{scalar}} v_j$ where $v_j$ is the velocity vector of the rotating scalar particles.   Using the change of variables defined above, $\Phi = \Sigma\, {\rm exp}(i\Gamma)$, we can use these expressions to find $v_j$:
\beq
\label{vj}
v_j = \frac{1}{m_H} \, (2eA_j-\partial_j \Gamma) \, .
\eeq
This known result is notably different from that of a superfluid in which the scalar field does not couple to a gauge field.  In the absence of the $A_j$ term in the above expression, one would conclude that ${\bf \nabla} \times {\bf v} = 0$ and thus the scalar condensate does not support rotation.  Instead, in the presence of the gauge field we find
\beq
\label{rotv}
{\bf \nabla} \times {\bf v} = \frac{2e}{m_H} \, {\bf H} \, .
\eeq 
The magnetic field ${\bf H}$ is called the London field \cite{London}.

The velocity vector ${\bf v}$ can be written in term of the angular velocity ${\bf v} = {\bf \Omega} \times {\bf r}$.  It follows that, for constant ${\bf \Omega}$, the rotation of ${\bf v}$ is given by ${\bf \nabla} \times {\bf v} = 2 {\bf \Omega}$.  Accordingly, the magnetic field can be expressed in terms of the angular velocity:
\beq
\label{HLondon}
{\bf H} = \frac{2m_H}{2 e} \, {\bf \Omega} =  \frac{2eJ_0}{m_\gamma^2} \, {\bf \Omega} \, .
\eeq
Here $J_0$ is the fermionic number density.  Thus the condensate of helium-4 nuclei can rotate with the rest of the star, unlike a neutral condensate.  The consequence is a small, constant magnetic field in the bulk of the condensate.

Varying the Lagrangians (\ref{Lgww1}) and (\ref{lagrF}) with respect to $A_j$ gives $2 e J_j ^{\rm{scalar}} = e J_j$, where $J_j$ is the fermion current density.  Using $J_j = J_0 v_j$, it follows that the fermion velocity vector is equal to the scalar velocity vector.  The electrons and the helium-4 nuclei rotate together in the core of the star.  At the surface however, there is a thin layer of helium-4 nuclei that is slightly out of rotation with the rest of the star.  This feature becomes evident upon finite volume regularization of the system.  The thickness of the layer is roughly $1/m_\gamma$.  This surface layer is what gives rise to the London field in the interior of the star \cite{Sauls}.  To estimate the angular velocity of a helium white dwarf star we take $\Omega \sim 10^{-2}$ Hz.  The resulting London field is $H \simeq 10^{-6}$ Gauss.  This field is present even in the absence of vortices.  However, it is too small to affect any of the results given above.

The existence of vortices in white dwarf stars with condensed cores could have observational consequences.  In particular, one could consider the scattering of light near the surface of the star.  Moreover, the discussions in the previous chapter concerning the specific heat of the star would be modified with the existence of vortices, and calculations of the cooling rate and the luminosity function would have to be adjusted accordingly.

\subsubsection*{Acknowledgments}

We are grateful to David Pirtskhalava for many valuable discussions and 
collaboration on the helium-core dwarf cooling. We'd like to thank
Nima Arkani-Hamed, Paul Chaikin, Alexander Dolgov,  
Daniel Eisenstein, Leonid Glazman, Andrei Gruzinov, 
Stefan Hofmann, Pierre Hohenberg, 
Juan Maldacena, Aditi Mitra, Slava Mukhanov, 
Hector Rubinstein, Malvin Ruderman, Dan Stein, Arkady Vainshtein, 
and Alex Vilenkin for useful discussions and  correspondence.
The work of GG was supported by NSF grant PHY-0758032; most of this work was done while   RAR was supported by James Arthur Graduate 
Fellowship at NYU.  RAR also acknowledges support from the Swedish Research Council (VR) through the Oskar Klein Centre.

\clearpage


\end{document}